\documentclass[fleqn,usenatbib]{mnras}

\usepackage[T1]{fontenc}
\usepackage{booktabs}
\usepackage{graphics}
\usepackage{caption}
\usepackage{subcaption}
\usepackage{physics}
\usepackage[demo]{graphicx}
\usepackage{cleveref}
\usepackage{multirow}

\DeclareRobustCommand{\VAN}[3]{#2}
\let\VANthebibliography\thebibliography
\def\thebibliography{\DeclareRobustCommand{\VAN}[3]{##3}\VANthebibliography}

\usepackage{graphicx}	
\usepackage{amsmath}	
\usepackage{amssymb}	
\usepackage{adjustbox}
\usepackage{newtxtext,newtxmath}

\title[Stellar Populations $\&$ ISM in central ETGs]{Probing Stellar Populations and Interstellar Medium in Early-Type Central Galaxies}

\author[Lorenzoni, V. et al.]{
 Vanessa Lorenzoni,$^{1}$\thanks{E-mail: vanessa.lorenzoni@acad.ufsm.br} 
 Sandro B. Rembold,$^{1}$
 Reinaldo R. de Carvalho$^{2}$
\\
$^{1}$Departamento de Física, CCNE, Universidade Federal de Santa Maria, 97105-900, Santa Maria, RS, Brazil\\
$^{2}$ NAT - Universidade Cidade de São Paulo, 01506-000, São Paulo, SP, Brazil
}

\pubyear{2022}

\begin{document}
\label{firstpage}
\pagerange{\pageref{firstpage}--\pageref{lastpage}}
\maketitle

\begin{abstract}
In this study, we analyse the characteristics of stellar populations and the interstellar medium (ISM) in 15,107 early-type central galaxies from the SPIDER survey. Using optical spectra from the Sloan Digital Sky Survey (SDSS), we investigate stellar age (Age), metallicity ($Z$), visual extinction ($A_{\rm V}$), and H$\alpha$ equivalent width (EWH$\alpha$) to understand the evolution of the baryonic content in these galaxies. Our analysis explores the relationship between these properties and central velocity dispersion ($\sigma$) and halo mass ($M_{\rm halo}$) for isolated centrals (ICs) and group centrals (GCs). Our results confirm that both ICs and GCs' stellar populations and gas properties are mainly influenced by $\sigma$, with $M_{\rm halo}$ playing a secondary role. Higher $\sigma$ values correspond to older, more metal-rich stellar populations in both ICs and GCs. Moreover, fixed $\sigma$ values we observe younger Ages at higher values of $M_{\rm halo}$, a consistent trend in both ICs and GCs.  Furthermore, we investigate the ionisation source of the warm gas and propose a scenario where the properties of ionised gas are shaped by a combination of cooling within the intra-cluster medium (ICM) and feedback from Active Galactic Nuclei (AGN) assuming a Bondi accretion regime. We observe inherent differences between ICs and GCs, suggesting that the ratio between AGN kinetic power and ICM thermal energy influences EWH$\alpha$ in ICs. Meanwhile, gas deposition in GCs appears to involve a more complex interplay beyond a singular AGN-ICM interaction.
\end{abstract}




\section{Introduction}

In thermalised clusters, a central galaxy, typically bright and massive, resides at the cluster's core. Even in dynamically unrelaxed clusters, the central galaxy can be identified as the one hosting the most massive dark matter sub-halo \citep{Yang11}. These central galaxies often display a spheroidal morphology and consist of old, metal-rich stars with little to no ongoing star formation \citep{Loubser}. They are commonly referred to as Brightest Cluster Galaxies (BCGs) and exhibit distinct physical properties compared to non-central early-type galaxies (ETGs) of similar mass. In fact, they can be up to ten times more luminous than typical ETGs and do not follow the luminosity function observed in other cluster members \citep{Katayama, Dressler}. Moreover, central galaxies residing in massive clusters tend to be associated with radio emission  \citep{Von, stott} and possess a cD morphological type \citep{dress, 2015Zhao, 2015Zhao2}. Consequently, it is reasonable to consider alternative formation pathways for these objects.

Environmental factors seem to contribute to the evolution of central galaxies. It has being observed \citep{2014ApJ...788...29L} that groups and clusters centrals exhibit distinct characteristics from isolated centrals (halos lacking satellites galaxies). Specifically, centrals in groups or clusters tend to display a redder colour, higher mass, and reduced star formation activity. Moreover, isolated centrals at a fixed $\sigma$ have older ages, higher [$\alpha$/Fe] ratios, and lower internal reddening compared to their counterparts in groups \citep{La_Barbera_2014}. These authors also propose that stars in central galaxies of more massive halos are younger, implying that group or cluster centrals formed their stellar components over longer timescales than isolated ones. In a separate investigation, \citet{2015MNRAS.453.4444Z} reveal that more massive BCGs predominantly reside in denser regions and more massive clusters when compared to lower mass BCGs.

The formation and evolution of central galaxies remains a complex and unresolved topic, involving internal processes like gas-to-star transformation and feedback from supernovae and AGN, as well as external processes associated with the cluster environment. 
An example of an environmentally-driven evolutionary process is the cooling flow observed in cool-core clusters. For instance, the ICM gas rapidly (within 10 to 100 million years) loses pressure support and thermal energy through X-ray emission \citep{2013AN....334..394G}, namely the cooling flow supplies cold gas to the central galaxy, which may induce star formation. \cite{1996ApJ...466L...9M} present clear evidence of this process as blue and ultraviolet excesses indicate star formation in the vicinity to the central galaxy. Cold molecular gas associated with the cooling flow and H$\alpha$-emitting filaments from young stars near the cluster core have also been observed  \citep{2001MNRAS.328..762E, 2004A&A...415L...1S, 2008A&A...477L..33R, 2006A&A...454..437S}. Moreover, \cite{2012Natur.488..349M} investigate a galaxy cluster at redshift $z = 0.596$ with an exceptionally strong cooling flow and observe a massive starburst in the central galaxy.
However, it is noteworthy that the cooling flow rates often fall below predictions from the classic model \citep{1989McNamara}, which suggests an intense star formation rate of $1,000\,\rm{{M}_{\odot}}\,\rm yr^{-1}$  \citep{1994Fabian}. Observations indicate that centrals exhibit star formation rates ranging from approximately 1\% to 10\% of the predicted value \citep{2001MNRAS.328..762E, Odea, McDonald10, McDonald2011b}, implying that the ICM is reheated by an external source  \citep{1994Fabian}.

Observational and simulation-based studies, such as  \cite{2006MNRAS.365...11C,2007ARA&A..45..117M} and \citet{2008ApJ...686..859B}, suggest that radio-mode feedback acts as a regulator of gas supply in cooling flows. The feedback mechanism produces powerful jet structures with radio-emitting lobes, observable as X-ray cavities causing depressions in surface brightness. The spatial correspondence between these structures and the radio jets are important signatures of the interplay between AGN feedback and the host environment \citep{2004ApJ...607..800B,2011ApJ...734L..28C,2015ApJ...805...35H}.
However, the mode and efficiency of the accretion mechanism responsible for producing radio jets as well as the energy required to regulate cooling flows are still debated. \citet{2006MNRAS.368L..67B,2006MNRAS.372...21A,2007MNRAS.376.1849H,2014arXiv1406.6366F} argue that jet formation is primarily driven by accretion of hot gas from the X-ray emitting medium, known as Bondi accretion \citep{1952MNRAS.112..195B}. On the other hand, \cite{2015ApJ...799L...1V} propose a model of precipitation-regulated AGN feedback in central galaxies, where gas from the ICM flows toward the cluster centre, fuelling the supermassive black hole (SMBH) through cold and chaotic accretion \citep{2013MNRAS.432.3401G}.  This scenario of jet formation via cold gas accretion is also supported by \citet{2011ApJ...727...39M}.
In both scenarios, the jets produced expel gas from the galaxy, leading to self-regulation between AGN feedback and the feeding of the SMBH. AGN feedback can not only regulate the cooling flow in cool-core clusters but it also impacts the properties of the central galaxy's baryonic content: it is crucial for reproducing the bright end of the galaxy luminosity function and explaining the red colours and chemical composition of central galaxies, indicating a short duration of the star formation epoch \citep{2005ApJ...620L..79S}.

Interactions between galaxies within clusters are also frequently observed, with dynamic friction \citep{1943ApJ....97..255C} playing a crucial role in the migration of massive galaxies towards the centre of galaxy clusters.  This process drives the merging and growth of central galaxies, while also triggering star formation and black hole activity.
Consequently, it provides a potential explanation for the observed high masses and the very extended radial surface brightness distributions found in cD galaxies \citep{Oemler}. \citet{2022Oyarz} suggest that mergers in central galaxies can trigger rapid and intense star formation episodes, leading to stellar mass growth. Central galaxies with higher galaxy total mass exhibit older stellar populations and appear to have formed over shorter time-scales than those with lower stellar mass. \citet{2022AJ....163..146S} and \citet{La_Barbera_2014} propose mergers with gas-rich galaxies as the primary fuel source for AGNs in high-redshift BCGs. Moreover, mergers with gas-rich galaxies potentially explain the higher metallicity and younger ages observed in group centrals compared to isolated centrals.

The assembly history of central galaxies varies with central galaxy mass and host halo, as indicated by previous studies \citep{La_Barbera_2014, 2022Oyarz}. Despite progress in understanding their formation and evolution, several questions remain unanswered. Specifically, the development of the stellar and gas content of central galaxies and the relative contributions of internal and external processes in the assembly history of such objects are still open questions.
In this study, we address these questions by investigating the stellar population properties of central galaxies as a function of their $\sigma$ (a proxy for total galaxy mass) and $M_{\rm halo}$ (a proxy for the environment). 
We conduct a comprehensive analysis of stellar populations and ionised gas properties in a sample of central galaxies spanning various ranges of $\sigma$ and $M_{\rm halo}$. We quantitatively investigate the dependencies of these parameters in both regimes, employing simple analytical functions for parameterisation. Furthermore, we examine the observed characteristics of ionised gas within a scenario that considers the interplay between feedback from the central galaxy's nuclear activity and intracluster gas cooling, assuming the Bondi accretion model, to regulate the gas accretion into the central galaxy.

The paper is structured as follows: Section \ref{sec:Data} provides details on the central galaxy sample selection, the stacking procedure and how we run the stellar population synthesis; In Section \ref{sec:Results}, we present our results, considering isolated and group centrals separately. In Section \ref{sec:discussion} we discuss the main findings of our investigation presenting a model of the gas content in central galaxies Finally, our conclusions are summarised in Section \ref{sec:Summary}.
Throughout the present work, we adopt $\Omega_m = 0.3$, $\Omega_{\Lambda} = 0.7$ and $H_0 = 100\,h\,\rm{km\,s^{-1}\,Mpc^{-1}}$ , with $h = 0.7$.

\section{Data and methods}
\label{sec:Data}

Our sample is based on Spheroid’s Panchromatic Investigation in Different Environmental Regions --  SPIDER I \citep{spideri}, which includes $39,993$ ETGs within the redshift range $0.05 < z < 0.095$ from the Sloan Digital Sky Survey Data Release 6 \citep[SDSS-DR6]{data}. We apply an absolute r-band Petrosian magnitude limit of $M_{r} < -20$, corresponding to the separation between \textit{bright} and \textit{ordinary} elliptical galaxies \citep{Capaccioli,Graham}, which aligns approximately with the upper redshift limit of $z = 0.095$, where SDSS spectroscopy is complete. The lower redshift limit avoids the aperture effect in our measurements.

For selecting ETGs, we adopt the criteria established by \citet{spideri}. We set the SDSS attribute $fracDev_r > 0.8$ to select bulge-dominated systems and the attribute $eClass < 0$ to determine spectral type based on principal component analysis decomposition. We also include only objects with $\sigma > 100\,\rm{km}\,\rm{s}^{-1}$, to remove the low-mass ETGs. We exclude galaxies with low internal extinction ($E(B-V) < 0.1\,\rm{mag}$) to remove contamination by lenticular galaxies, as well as spectra with signal-to-noise ratio (S/N, per \AA) in the $\rm H_{\beta}$ region smaller than $14$, $27$, and $21$ for $\sigma = 100$, $200$, and $300\,\rm{km}\, \rm{s}^{-1}$ respectively; such S/N cuts correspond to the lowest S/N quartile in the given $\sigma$ bin. We also verify the morphological classification of these galaxies using the Galaxy Zoo project \citep{Lintott}. However, given that not all galaxies have Galaxy Zoo classifications, we apply additional selection criteria based on the quality of the two-dimensional Sérsic fits to the surface brightness distribution, as described in \citet{La_Barbera_2014}. In the end, we are left with $20,977$ ETGs with optical spectra, $\sigma$ and its uncertainty ($\sigma_{\rm err}$) extracted from SDSS-DR12 \citep{Alam}. The uncertainties in $\sigma$ are typically less than $10\,\rm km\,s^{-1}$, except for 95 central galaxies (with the maximum $\sigma_{\rm err}$ value being $13.68$), a negligible figure considering all central galaxies in our sample.

\subsection{Characterisation of the galaxy host environment}
\label{EnvCharacterisation}

The galaxy host environment is characterised by the Yang Catalogue \citep[hereafter Y07]{yang}. Y07 is constructed using a halo mass finder algorithm applied to the New York University Value-Added Galaxy Catalogue \citep[NYU-VAGC]{2005AJ....129.2562B}. Galaxies within a common dark matter (DM) halo form a group and are classified as either satellite or central galaxies, based on their stellar mass.

Central galaxies are identified as the most massive galaxy within the DM halo and can be distinguished in two categories: 1) isolated centrals (IC), those occupying a single DM halo and no other galaxy; 2) group centrals (GC), those sharing a DM halo with other satellite galaxies. Distinct evolutionary processes may shape the physical properties of ICs and GCs differently. Therefore, we investigate these two populations separately. Y07 determine the DM halo mass by the total stellar mass of the group, which must satisfy the condition $M_{\rm halo} \gtrsim 10^{12}\,$h$^{-1}\,\rm{M_{\odot}}$. Our sample includes $15,107$ early-type central galaxies, where $10,575$ are ICs and $4,532$ are GCs.

\begin{figure}
    \includegraphics[width=\columnwidth]{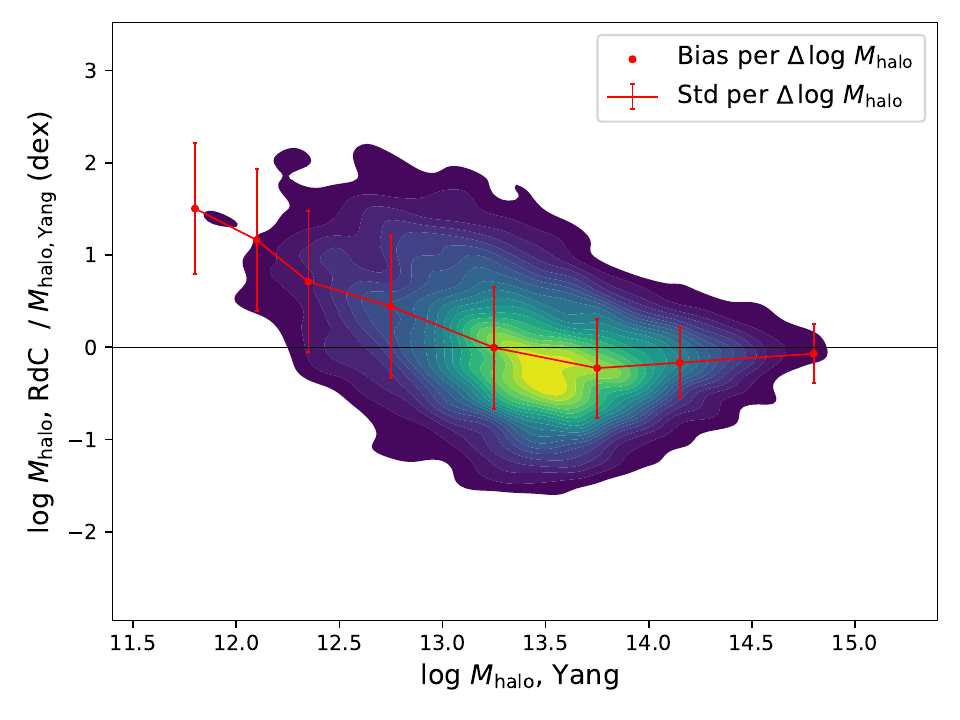}
    \caption{Comparison between the halo mass measurements obtained by \citet{2017AJ....154...96D} and by \citet{yang}. The red dots and lines indicate the bias and standard deviation for specific halo mass ranges, respectively.}
    \label{gapper}
\end{figure}

We compare halo masses obtained using the shift-gapper technique by \citet{2017AJ....154...96D} with those estimated by \cite{yang} to evaluate systematic errors in $M_{\rm halo}$ measurements. The comparison, shown in Figure \ref{gapper}, is based on the dispersion and bias generated by these two methods for specific ranges of the Yang halo mass ($\Delta\,\log M_{\rm halo})$. 

\subsection{Stacking galaxy spectra by their $\sigma$ and $M_{\rm halo}$}
\label{stacking_section}

To enhance the S/N of optical spectra, we perform spectral stacking by median-combining the flux-normalised individual spectra within bins of $\sigma$ and $M_{\rm halo}$. The resulting stacked spectra, corrected for redshift, is computed as the median of the available spectra within each bin. The flux uncertainties, for a given wavelength, are estimated from the standard error of the median across all spectra entering each final stacked spectrum. Thus, our sample is distributed across $71$ stacks, with $\sigma$ ranging from $100$ to $323$\,km\,s$^{-1}$ and $M_{\rm halo}$ spanning from $10^{11.6}$ to $10^{15.3}\,\rm{M_{\odot}}$. The bin size is set to vary in steps of $10\,\rm{km\,s}^{-1}$ for $\sigma$, except for the highest-value bins. 
For $M_{\rm halo}$, we vary the bin width in steps of $0.25$, $0.40$, $0.50$, and $1.00$\,dex. The initial range for $M_{\rm halo}$ begins at $10^{11.6}\,\rm{M_\odot}$, as our sample does not include any halos below this threshold.
Finally, we exclude from further analysis all stacks containing less than 5 galaxies, thus ensuring a minimum representative number of objects per stack. The number of galaxies and the S/N per bin are summarised in Tables~\ref{ngal}, \ref{tab:ngal_isolated} and \ref{tab:ngal_group}. The ICs, which predominantly occupy halos with masses below $10^{13.5}\,\rm{M_{\odot}}$, are concentrated in 54 bins. In contrast, GCs are predominantly associated with higher $M_{\rm halo}$ values, and yield 61 bins of GC galaxies. This highlights the connection between galaxy properties and DM halos, as reflected in the stellar-to-halo mass relation (SHMR), where more massive galaxies are typically found in more massive halos.
It is important to note that the stacking process enhances the spectral S/N, but removes individual signatures from each galaxy, providing a representative spectrum of the entire population.

\subsection{Stellar population properties: Spectral synthesis analysis using {\sc starlight} code}
\label{SPS} 

To estimate Age, $Z$ and $A_{\rm V}$, we use the spectral synthesis method by employing the {\sc starlight} software \citep{Cid05}, which performs spectral synthesis by comparing the observed galaxy spectrum with a set of simple stellar population (SSP) models, searching for the combination of SSPs that best represents the observed spectrum. In this study, we use a set of 108 solar-scaled SSPs models from the Medium resolution INT Library of Empirical Spectra (MILES) galaxy spectral library \citep{Miles}. These models are constructed using the Kroupa universal Initial Mass Function (IMF) and cover ages ranging from $0.5$ to $17.78\,$Gyr,  with $Z = 0.004$, $0.008$, $0.019$ and $0.03$. The possible occurrence of non-solar abundance ratios in our sample of galaxies is not expected to affect significantly the derived stellar population parameters, as discussed in \citet{La_Barbera_2014}.

The MILES models have a spectral resolution of approximately $2.5\,$\AA\ and cover a wavelength range from $3,525$ to $7,500\,$\AA. In line with the study by \citet{La_Barbera_2014}, we conduct the synthesis using the wavelength range from $4,000$ to $5,700\,$\AA, with $5,200\,$\AA\ as the normalisation wavelength; the exclusion of the spectral range above $5,700\,$\AA{} is motivated by the presence of particularly IMF-sensitive absorption features (like Na and TiO) in this region. We apply the \citet{CCM} extinction law (suitable for systems with low levels of star formation) and mask the main SDSS optical emission lines. We estimate the light-weighted Age and $Z$ following \citet[, their Equations 2 and 5]{Cid05}, while the $A_{\rm V}$ parameter is directly provided by the {\sc starlight} output.

To ensure the robustness of our results, we perform additional syntheses using an alternative set of SSPs, the Granada-MILES (GM) models \citep{CidGM}. GM templates are a combination of Granada models from \citet{Granada} (for ages younger than $63\,$Myr) and MILES models. The construction of the SSPs uses a Salpeter IMF and includes the evolutionary tracks by \citet{2000A&AS..141..371G}, with the exception of ages $\leq 3\,$Myr, where the Geneva evolutionary tracks are employed \citep{1992A&AS...96..269S,1993A&AS...98..523S,1993A&AS..102..339S,1993A&AS..101..415C} and a spectral resolution of approximately $2.3\,$\AA. We convolve the GM SSPs spectra to coincide the spectral resolution of the MILES models and also select only the templates with the same age and $Z$, enabling a direct comparison between them. We have found  (Figure ~\ref{fig:mh_sigma_GM}) that differences in SSPs can indeed affect the derived stellar population parameters; however, the derived EWH$\alpha$ values are entirely insensitive to the chosen SSPs, and all stellar population parameter trends with $\sigma$ and $M_{\rm halo}$ are consistent. We are therefore confident that the observed $\sigma$ and $M_{\rm halo}$ trends that are presented and discussed in Sect.~\ref{sec:Results} are not dependent on the choice of SSPs models.

To estimate the equivalent width of H$\alpha$ (EWH$\alpha$), we extend the synthesis solution obtained previously up to $7,000\,$\AA\ to include H$\alpha$ emission. We then subtract the synthetic and observed spectra to obtain the residuals, which contain only the contribution of emission lines. Finally, the H$\alpha$ flux (Gaussian area) is divided by the mean level of the stellar continuum at the line position, obtained from the synthetic spectrum. The only purpose of this second synthesis run up to $7,000\,$\AA\ is to allow for a straightforward measurement of EWH$\alpha$; the stellar population parameters that will be presented and discussed in the remainder of the paper are those obtained with the synthesis limited to $5,700\,$\AA.

Uncertainties in the studied properties are estimated using the bootstrap method. In short,  we perform a random selection of a single object among all galaxies in a given stack; this operation is repeated $N$ times, where $N$ is the total number of galaxies in the respective stack. In this process, not all galaxies of the stack are selected, because a given galaxy can be selected multiple times. Subsequently, the entire stack creation process is performed with the newly selected galaxies. Then, we conduct the stellar population synthesis and estimate each property mentioned before. This procedure is repeated 1,000 times for each stack. To obtain the uncertainty in the parameters, we calculate the standard deviation for each stack and property.

\section{Results}
\label{sec:Results}

 \begin{figure*}
	\includegraphics[width=16cm]{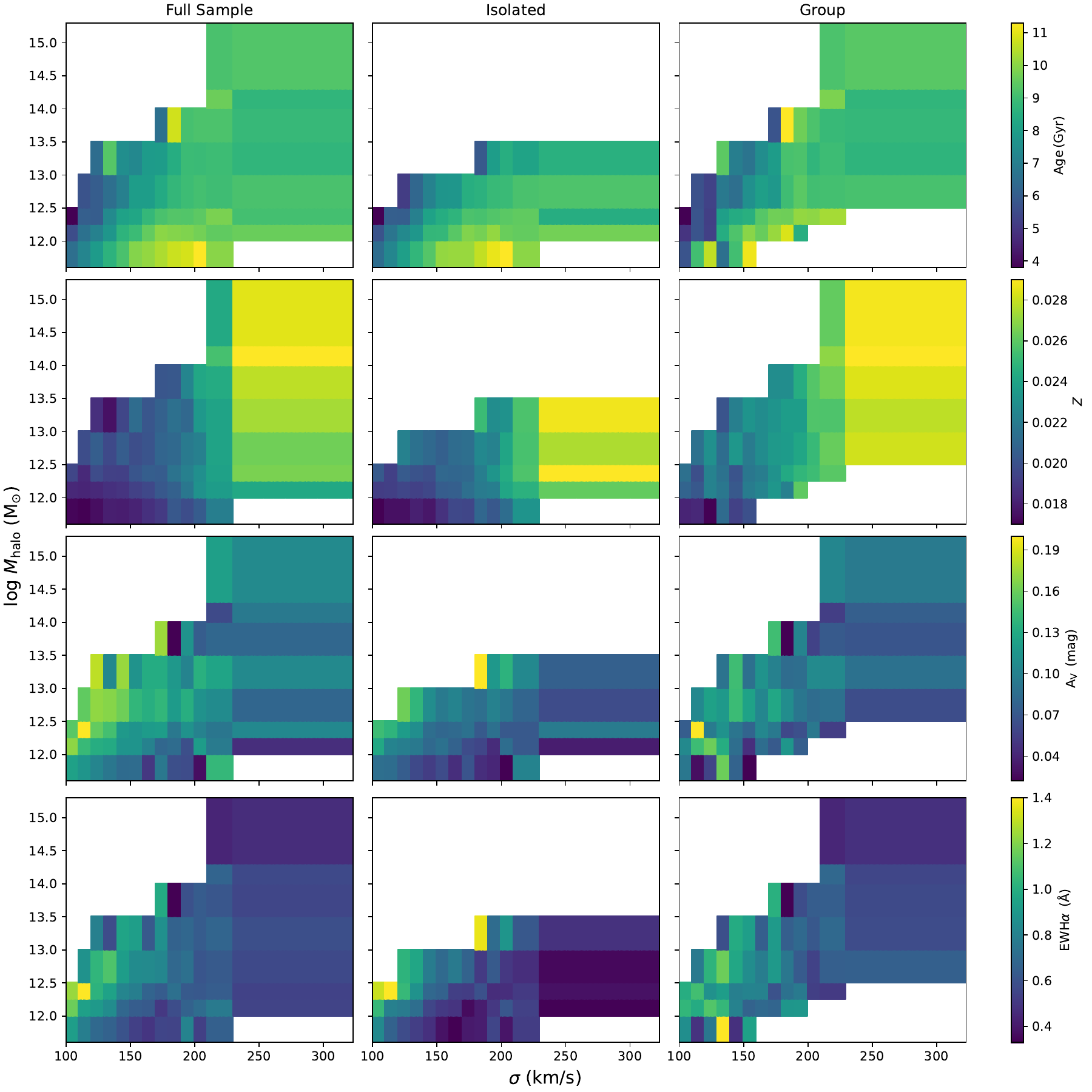}
    \caption{Variations in Age, $Z$, $A_{\rm V}$, and EWH$\alpha$ as a function of $\sigma$ (horizontal axis) and $M_{\rm halo}$ (vertical axis) are shown from the top to bottom, respectively. The full sample, as well as ICs and GCs, are illustrated from left to right, respectively. The physical properties are represented by colours.}
    \label{fig:resvaz}
\end{figure*}

\begin{figure*}
	\includegraphics[width=\textwidth]{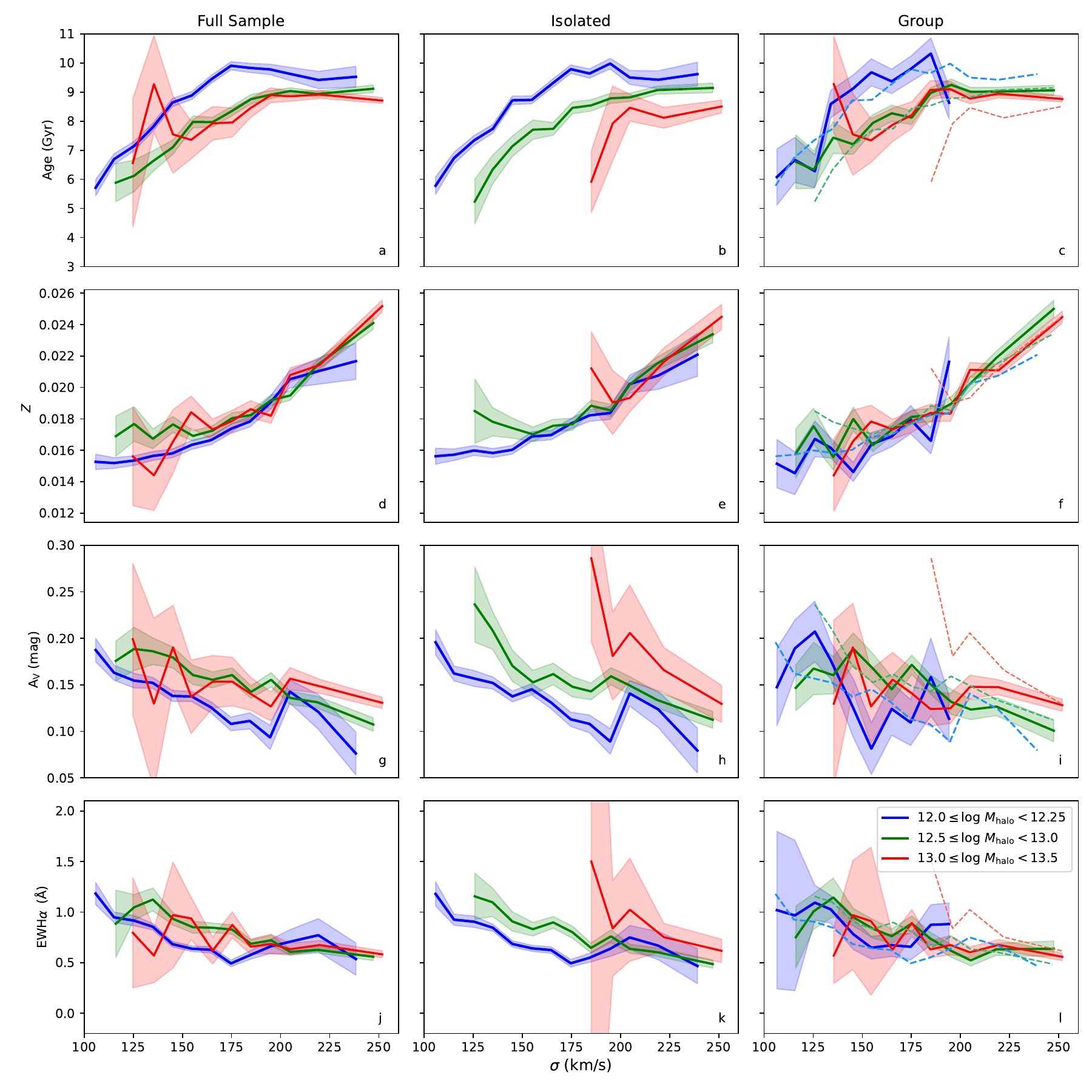}
    \caption{Relationship between Age, $Z$, $A_{\rm V}$, and EWH$\alpha$ with $\sigma$, depicted from top to bottom, respectively. The full sample, along with ICs and GCs, is presented from left to right, respectively. Additionally, we included three $M_{\rm halo}$ ranges, distinguished by colours.}
    \label{fig:mh_sigma}
\end{figure*}

\subsection{Exploring the trends in stellar population and interstellar medium properties}
\label{overall_trends}

In Figure~\ref{fig:resvaz}, we provide an illustrative overview of all parameters, emphasising their inherent dependencies on both $\sigma$ and $M_{\rm halo}$ , despite the presence of significant scatter. To further explore these dependencies, we narrow our focus to specific $M_{\rm halo}$ ranges. As shown in Figure~\ref{fig:mh_sigma}, we examine the behaviour of each observable as a function of $\sigma$ for both ICs and GCs, employing three distinct $M_{\rm halo}$ bins, distinguished by different colours, to simplify the visualisation of systematic trends. These $M_{\rm halo}$ ranges were chosen for their availability to both ICs and GCs. Also, we have included the curves obtained for ICs as dotted lines in the GCs panels, making a direct comparison easier.


\subsubsection{Stellar Age - $\rm Age$}
\label{Trends_age}

As far as ICs are concerned, we see that Age increases up to approximately $\sigma \sim 200\,\rm km\,s^{-1}$ and then reaches a plateau (panel b of Figure~\ref{fig:mh_sigma}). Higher values of $M_{\rm halo}$ are associated with older stellar populations at a given $\sigma$. However, for  $\sigma$ values lower than $150\,\rm{km\,s^{-1}}$, the large uncertainties make it more difficult to interpret the trend. Notably, there is an age difference of approximately $2\,$Gyr between the low and high $M_{\rm halo}$ bins at a given $\sigma$. These trends are basically the same as those exhibited by GCs (panel c of Figure~\ref{fig:mh_sigma}), although with a greater scatter due to the lower number of galaxies per bin. It is worth noticing that there are no GCs with $\sigma \geq 200\,\rm km\,s^{-1}$ in $12.0 \leq \log\,M_{\rm halo} \leq 12.25$. Despite these findings being consistent with those presented by \citet{La_Barbera_2014}, we do not observe any discernible difference between ICs and GCs, contrary to their discussion. This crucial distinction will be further addressed in Sect.~\ref{sec4.1}.

\subsubsection{Stellar Metallicity - $Z$}
\label{Trends_met}

We note a tendency for the stellar population to be more metal-rich as $\sigma$ increases, the so called mass-metallicity relation; this holds for ICs and GCs indistinctly (panels e and f of Figure~\ref{fig:mh_sigma}). We do not find any dependence on $M_{\rm halo}$, as indicated by the overlapping curves, except for centrals with $\sigma  \lesssim 150\,\rm{km}\,\rm{s}^{-1}$, for which we note a slight increase in $Z$ for the intermediate $M_{\rm halo}$ bin. This result is in agreement with those reported by \citet{La_Barbera_2014}. There seems to be no difference between ICs and GCs as far as the behaviour of $Z$ with $\sigma$ is concerned, albeit with greater scatter for GCs.

\subsubsection{Visual Extinction - $A_{\rm V}$}
\label{Trends_av}

Examining panels h and i of Figure~\ref{fig:mh_sigma} we identify a decrease in $A_{\rm V}$ as $\sigma$ increases, for ICs and GCs. Particular, for ICs we find a systematic increase in $A_{\rm V}$ as we probe centrals with $\sigma \lesssim 200\,\rm km\,s^{-1}$ and greater $M_{\rm halo}$. For $\sigma \gtrsim 200\,\rm km\,s^{-1}$ the curves overlap and $A_{\rm V}$ becomes insensitive to $M_{\rm halo}$. Conversely, there appears to be no correlation between $A_{\rm V}$ and $M_{\rm halo}$ for GCs. Nevertheless, the scatter is significantly large, making it inconclusive to assert that the same systematic difference with $M_{\rm halo}$ observed in ICs is also present in GCs.

\subsubsection{Equivalent width of H$\alpha$ - $\rm EWH\alpha$}
\label{Trends_ew}

In line with $A_{\rm V}$, ICs exhibit a decreasing trend in EWH$\alpha$ as $\sigma$ increases. Notably, we observe a systematic variation with $M_{\rm halo}$, with centrals with higher EWH$\alpha$ inhabiting more massive halos. The difference in EWH$\alpha$ amounts to approximately $0.6$\,\AA, which is significantly larger than the errors in EWH$\alpha$ for the two lowest $M_{\rm halo}$ bins in the plot. The relationship between EWH$\alpha$ and $\sigma$ is less pronounced for GCs. Moreover, no apparent correlation between EWH$\alpha$ and $M_{\rm halo}$ is observed for GCs. The similarity in the results for $A_{\rm V}$ and EWH$\alpha$ holds true for both ICs and GCs, and further details will be discussed in Sect.~\ref{ICMAGN}.

\subsection{Correlating central galaxies properties with $\sigma$ and $M_{\rm halo}$}
\label{Parameterisation_with_sigma_and_halo_mass}

In the previous section, we provided a qualitative description of the studied properties, Age, $Z$, $A_{\rm V}$ and EWH$\alpha$, as a function of $\sigma$ and $M_{\rm halo}$, revealing a dependence on these parameters (Figures~\ref{fig:resvaz} and \ref{fig:mh_sigma}). Consequently, we conduct a quantitative analysis of these dependencies assuming a linear combination of $\sigma$ and $M_{\rm halo}$ to represent the parameters describing the stellar and gas content:  

\begin{equation}
\label{eq:parameterisation}
 Y \sim \sigma^{A}M_{\rm halo}^{\textit{B}},
\end{equation}
where $A$ and $B$ are free parameters and $Y$ represents the quantity under study. 

The choice of a power law is motivated only by its simplicity. Hence, we conduct a least squares fit to the function $\log Y = A \log \sigma + B \log M_{\rm halo}$ $ +\ C$, with uncertainties in $Y$ given by standard deviation, estimated via bootstrap (Sect.~\ref{SPS}).

To incorporate uncertainties in $\sigma$ and $M_{\rm halo}$, we employ the bootstrap method, performing $1,000$ fits. The $\sigma$ uncertainty ($\sigma_{\rm unc}$) for each bin is determined through error propagation on the mean $\sigma$ values, considering the 95th percentile of $\sigma_{\rm err}$. For each bin and fit iteration, we generate a new $\sigma$ value from a standard normal distribution centred around the mean value of the $\sigma$ range, with an uncertainty of $\sigma_{\rm unc}$.

In the case of $M_{\rm halo}$, to prevent overlap resulting from the narrow gap between $M_{\rm halo}$ ranges, we assign a random value between zero and $1.5$ (the maximum absolute bias between the two methods, obtained from Fig.~\ref{gapper}), to each $M_{\rm halo}$ measurement, allowing for positive and negative variations.  
In each fit iteration, we calculate the new $M_{\rm halo}$ by multiplying the random value by the corresponding bias for each $M_{\rm halo}$ range and adding it to the mean value of $M_{\rm halo}$ in the bin. This approach ensures a consistent directional shift across all $M_{\rm halo}$ ranges.
Furthermore, we consider random errors by applying a random distribution with galaxy-to-galaxy $\sigma_{\rm err}$ to the new $M_{\rm halo}$ value.

The best-fit $A$ and $B$ coefficients were obtained by computing the weighted average of the coefficients obtained from the $1,000$ bootstraps. The uncertainties are estimated by determining the standard deviation of the $1,000$ coefficients, weighted according to their respective errors.

Regarding the Age parameter specifically, we have observed that the best-fit coefficients grossly overestimate the ages of central galaxies at low $\sigma$ values. This is due to a large spread in the mean stellar age across the different bootstrap realisations, possibly related to a larger variety in stellar ages of the galaxies in such stacks. At high $\sigma$, we observe a much lower spread in Age, suggesting that galaxies at this $\sigma$ range are less varied regarding this parameter. Using the spread in Age as an estimate of its uncertainty results in an unsatisfactory parameterisation, since older central galaxy bins, characterised by lower spreads, influence disproportionately the solution, causing it to perform poorly for younger ages. Consequently, in order to find coefficients that describe are as accurately as possible the overall trends of all stacks, we ignore the Age uncertainties in the fits.

The values of the coefficients  $A$, $B$ and $C$, along with their corresponding uncertainties, are shown in Table~\ref{tab:coef_table} and confirm the trends observed in Figure~\ref{fig:mh_sigma}. The coefficient $C$ will no longer be considered, as it works only as a normalisation parameter. For instance, when examining the coefficients $A$ for Age, it becomes apparent that GCs exhibit a weaker correlation with $\sigma$ compared to ICs. In terms of the relationship with $M_{\rm halo}$, ICs demonstrate a value approximately twice as high as GCs. On the other hand, GCs display a stronger correlation between $Z$ and $\sigma$ compared to ICs. Conversely, the association between $Z$ and $M_{\rm halo}$ is less pronounced for GCs. It is noteworthy that the coefficients $B$ are small, indicating a weak dependence of $Z$ on $M_{\rm halo}$. The correlation coefficients between $A_{\rm V}$ and $\sigma$ are comparable for ICs and GCs, considering the uncertainties. However, when examining the correlation between $A_{\rm V}$ and $M_{\rm halo}$, ICs show a significantly stronger relation compared to GCs, where the correlation is very weak. Regarding the correlation between EWH$\alpha$ and $\sigma$, ICs and GCs display a similar level of correlation, although slightly weaker for GCs. On the other hand, the correlation coefficients between EWH$\alpha$ and $M_{\rm halo}$ are significantly smaller for GCs, being approximately ten times smaller than those for ICs. It is important to note that the correlation patterns between $A_{\rm V}$ and EWH$\alpha$ observed in Sect.~\ref{overall_trends} are also evident in the coefficients $A$ and $B$, considering the uncertainties, for both ICs and GCs. Therefore, we highlight that the correlation between all properties, regardless of the sample, shows a stronger association with $\sigma$ than with $M_{\rm halo}$.

\begin{figure*}
    \centering
    \includegraphics[width=\textwidth]{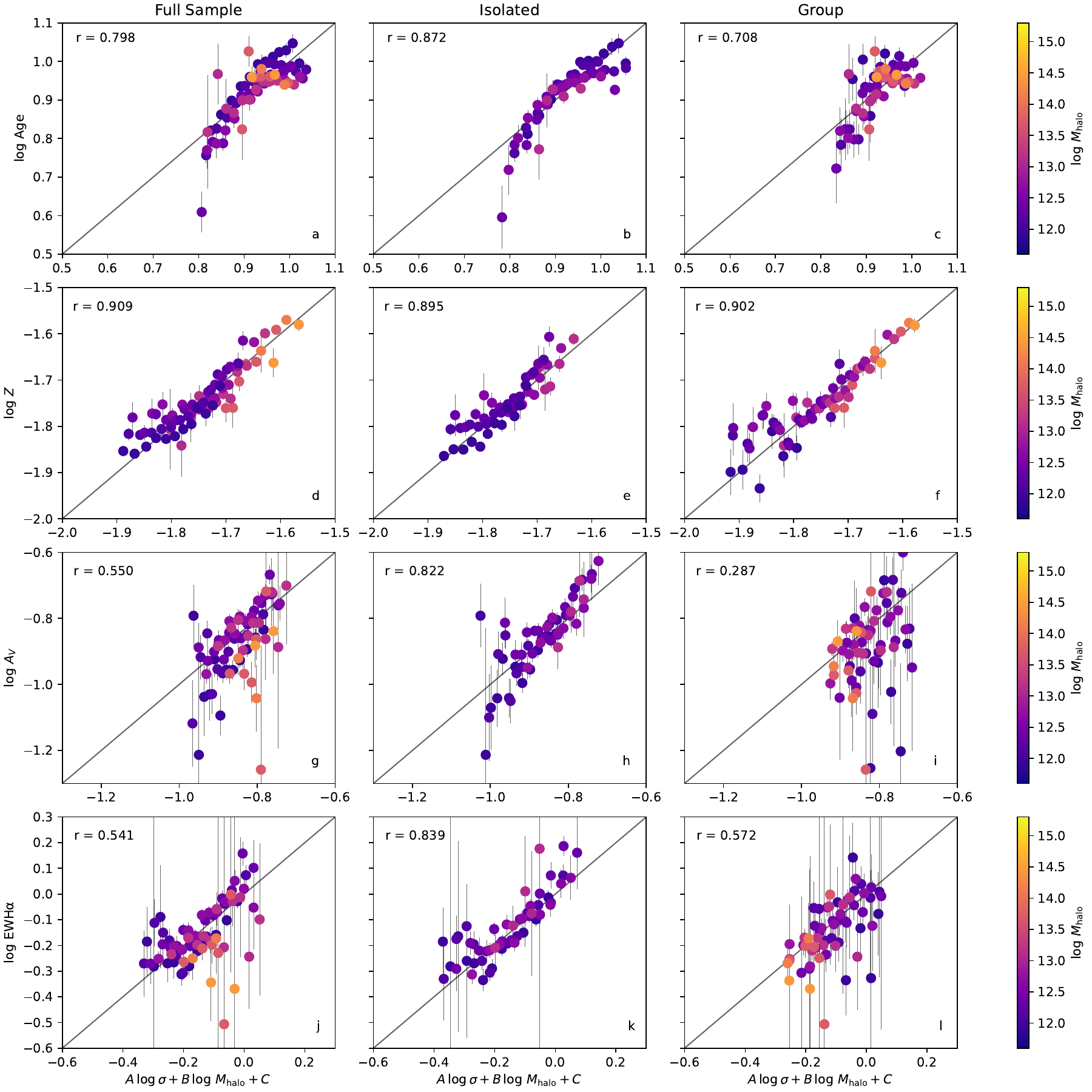}
    \caption{Comparison between observed properties and their corresponding predicted values obtained from Equation~\ref{eq:parameterisation}. The sequence of panels follows same pattern of Figure~\ref{fig:resvaz}. Central galaxies are represented by dots, with their corresponding $M_{\rm halo}$ ranges indicated by colours. Vertical error bars are derived from the bootstrap process detailed in Sect.~\ref{SPS}. Additionally, the degree of correlation between the observed and predicted values, estimated by the Pearson coefficient $r$, is shown in all panels.}
    \label{fig:comparison_A}
\end{figure*}

To assess the quality of the parameterisation, we present Figure~\ref{fig:comparison_A}, where we plot the observed properties against the predicted values obtained from Eq.~\ref{eq:parameterisation} using the best-fit coefficients from Table~\ref{tab:coef_table}. The panels of Figure~\ref{fig:comparison_A} follow the same arrangement as Figure~\ref{fig:mh_sigma}, presenting Age, $Z$, $A_{\rm V}$, and EWH$\alpha$ from top to bottom, respectively. From left to right, the panels display the results for the full sample, ICs, and GCs. The dots represent the physical properties values, with their corresponding $M_{\rm halo}$ indicated by colours. Vertical error bars are included to represent the standard deviation calculated using the bootstrap process described in Sect.~\ref{SPS}. We provide the Pearson coefficient ($r$) as a measure of the correlation between the observed and predicted values. Upon examining the panels b and c, a curved relationship is noticed, particularly for ICs, where the predicted values tend to overestimate the Age for $\log \rm{Age} < 0.8$. This non-linearity suggests that the functional form presented in Eq.~\ref{eq:parameterisation} may be inadequate to fully describe the relationship between Age, $\sigma$, and $M_{\rm halo}$. While this curvature consistently appears for central galaxies across different $M_{\rm halo}$, the primary source of discrepancy between the observed and predicted values is attributed to $\sigma$. It seems that the Age follows a relation with $\sigma$ that is less steep than a simple power law. Nevertheless, considering the relatively high Pearson coefficients ($r = 0.798$, $0.872$, and $0.708$ for the full sample, ICs, and GCs, respectively), these parameterisations capture most of the physical links between Age, $\sigma$, and $M_{\rm halo}$. However, it is important to interpret these coefficients with caution. The parameterisation for $Z$ demonstrates exceptional performance across all sub-samples. The scatter plot comparing the observed and predicted values of $Z$ closely follows a 1:1 relation, as evidenced by the high Pearson coefficients ($r = 0.909, 0.895, 0.902$ for the full sample, ICs, and GCs, respectively). Although Eq.~\ref{eq:parameterisation} effectively captures the relationship between $Z$ and both $\sigma$ and $M_{\rm halo}$, it is important to note that this is primarily due to the well known Mass-Metallicity relation and that $M_{\rm halo}$ contributes a small fraction to the final correlation. For $A_{\rm V}$, we achieve a good agreement between the observed and predicted values in ICs. Pearson's coefficient remains significant, with a value of $r = 0.822$ (panel h). However, the low Pearson coefficient of $0.287$ for GCs (panel i) indicates a predominance of dispersion in the data points, revealing a limitation in our parameterisation to accurately describe the relationship between $A_{\rm V}$, $\sigma$ and $M_{\rm halo}$. Similarly, for EWH$\alpha$, we also find a satisfactory agreement between the observed and predicted values for ICs, as indicated by the higher Pearson coefficient of $r = 0.839$ (panel k). However, for GCs, the scatter becomes more pronounced, resulting in a Pearson coefficient of $0.572$ (panel l). Despite the differences between the predicted and observed values being influenced by scattering for GCs, we consider the parameterisation to be a reasonable description for both $A_{\rm V}$ and EWH$\alpha$.

In a nutshell, the main findings of the parameterisation presented in Table~\ref{tab:coef_table} emphasise that the parameter $\sigma$ plays a more significant role in describing the observed properties of central galaxies compared to $M_{\rm halo}$. The influence of $M_{\rm halo}$ is generally small, especially for GCs. Also, ICs exhibit stronger correlations between most parameters and both $\sigma$ and $M_{\rm halo}$. However, there is an exception in the case of the correlation between $Z$ and $\sigma$, which is stronger for GCs. Therefore, predicting the behaviour of these quantities with respect to $\sigma$ and $M_{\rm halo}$ is more straightforward for ICs than for GCs. The implications of these findings will be discussed in the upcoming section.

\begin{table*}
\centering
\caption{Correlation coefficients $A$, $B$ and $C$ (column 1) obtained from Equation~\ref{eq:parameterisation}, for each sub-sample (column 2), between Age (column 3), $Z$ (column 4), $A_{\rm V}$ (column 5), and EWH$\alpha$ (column 6), versus $\log \sigma$ and $\log M_{\rm halo}$. The errors are quoted at the 1$\sigma$ level.}

\label{tab:coef_table}
\vspace{0.5cm}
\begin{tabular}{cccccc}
\hline
Coefficient & Environment  & Age & $Z$ & $A_{\rm V}$ & EWH$\alpha$ \\ \hline
\multirow{3}{*}{\textit{A}} & Full Sample     &  0.62 $\pm$ 0.03   &  0.58 $\pm$  0.04   &  -0.54 $\pm$ 0.15  &  -0.91 $\pm$ 0.18    \\  
                   & Isolated     &  0.70 $\pm$ 0.05   &  0.49 $\pm$  0.04   &  -0.62 $\pm$ 0.12  &   -1.14 $\pm$ 0.15   \\  
                   & Group &   0.53 $\pm$ 0.05   &  0.75 $\pm$  0.01   &   -0.55 $\pm$ 0.05  &  -0.82 $\pm$ 0.03    \\ \hline
\multirow{3}{*}{\textit{B} } & Full Sample     &   -0.033 $\pm$ 0.008  &  0.027 $\pm$  0.008   & 0.07 $\pm$ 0.02   &  0.09 $\pm$ 0.03    \\  
                   & Isolated     & -0.08 $\pm$ 0.04   &  0.03 $\pm$  0.02   &  0.13 $\pm$ 0.07  &  0.16 $\pm$ 0.08    \\  
                   & Group &  -0.033 $\pm$ 0.008   &  0.012 $\pm$ 0.004   &  0.016 $\pm$ 0.011  & 0.011 $\pm$ 0.007     \\ \hline
\multirow{3}{*}{\textit{C}} & Full Sample     &  -0.035 $\pm$ 0.13   &  -3.35 $\pm$ 0.14   &  -0.34 $\pm$ 0.33  &  0.95 $\pm$ 0.45    \\  
                   & Isolated     &   0.38 $\pm$ 0.55  &  -3.21 $\pm$ 0.24   &  -1.00 $\pm$ 0.82  &   0.56 $\pm$ 0.98   \\  
                   & Group &  0.15 $\pm$ 0.14  &  -3.59 $\pm$ 0.07   &  0.27 $\pm$ 0.09  &   1.60 $\pm$ 0.08   \\ \hline
\end{tabular}
\end{table*}

\section{Discussion}
\label{sec:discussion}
Understanding the evolution of central galaxies is a complex task, as it is influenced by many factors. The mass of the halos they inhabit plays a crucial role in determining their evolutionary trajectory. Also, $\sigma$ is a vital parameter when studying their stellar population and ionised gas content. A study conducted by \citet{La_Barbera_2014} focused on early-type central galaxies find that as $\sigma$ increases, these systems tend to be older. Furthermore, within a given $\sigma$ range, central galaxies in more massive halos appear to be younger. Adding to the intricacy of the situation is the interaction between the thermal energy of the intracluster medium and the kinetic power of the AGN, which introduces significant complexity to an already unresolved problem.

\subsection{The Impact of the Environment on the Star Formation History of Central Galaxies}
\label{sec4.1}

When studying how the environment influences galaxy properties, it is important to define the concept of ``environment'' carefully. \citet{La_Barbera_2014} use $M_{\rm halo}$ as an indicator of environment and consider a central galaxy as isolated when the $\log M_{\rm halo} \leq 12.5$. However, using Yang's classification of each structure, we observe contamination between ICs and GCs when employing La Barbera's definition of an IC. Specifically, there are 2011 ICs ($\sim 15\%$) with $\log M_{\rm halo} \geq 12.5$ and 889 GCs ($\sim 24\%$) with $\log M_{\rm halo} \leq 12.5$. In this study, we utilise Yang's structure classification to explore potential differences in the stellar populations of ICs and GCs. \citet{La_Barbera_2014} suggest that the difference between ICs and GCs stems mainly from ICs being defined as centrals with $\log M_{\rm halo} \leq 12.5$. However, considering the contamination mentioned earlier, this assertion appears questionable. The trend with $\sigma$ and $M_{\rm halo}$ is consistent for both ICs and GCs. This implies that regardless of whether the central galaxy is replenished with gas from its own halo (in the case of ICs) or from gas-rich systems that have been accreted (in the case of GCs), the final star formation history (SFH)  remains the same. It is crucial to remember that such a trend between Age and $\sigma$ is observed for galaxies in general (e.g., \citet{2015A&A...581A.103G}). Regarding the $Z$, panels e and f of Figure~\ref{fig:mh_sigma} yield results consistent with \citet{La_Barbera_2014}. Specifically, we find that $Z$ increases with $\sigma$, following the Mass-Metallicity relation initially established by \citet{1979A&A....80..155L}. Furthermore, in agreement with \citet{La_Barbera_2014}, we do not observe any dependence on $M_{\rm halo}$, except for a small difference between ICs and GCs when $\sigma \leq 135\,\rm{km}\,\rm{s}^{-1}$, with ICs being slightly more metal-rich than their counterparts in GCs.

The relationship between Age, $Z$, and $M_{\rm halo}$ has been a topic of debate in the literature. In a study by \citet{2022Oyarz} focusing on passive central galaxies from the MaNGA survey \citep{2015ApJ...798....7B}, it is observed that central galaxies are older and more metal-poor in more massive halos. The discrepancy between their findings and ours could be attributed to their selection of passive galaxies based on their spatially integrated specific star formation rate, regardless of their morphological type. In contrast, \citet{2022Scholz} find that, at a fixed $\sigma$, central galaxies tend to be more metal-poor in more massive halos. This difference may arise because our sample of central galaxies consists exclusively of ETGs, while the authors include both early- and late-type galaxies in their analysis.

\begin{figure*}
	\includegraphics[width=\textwidth]{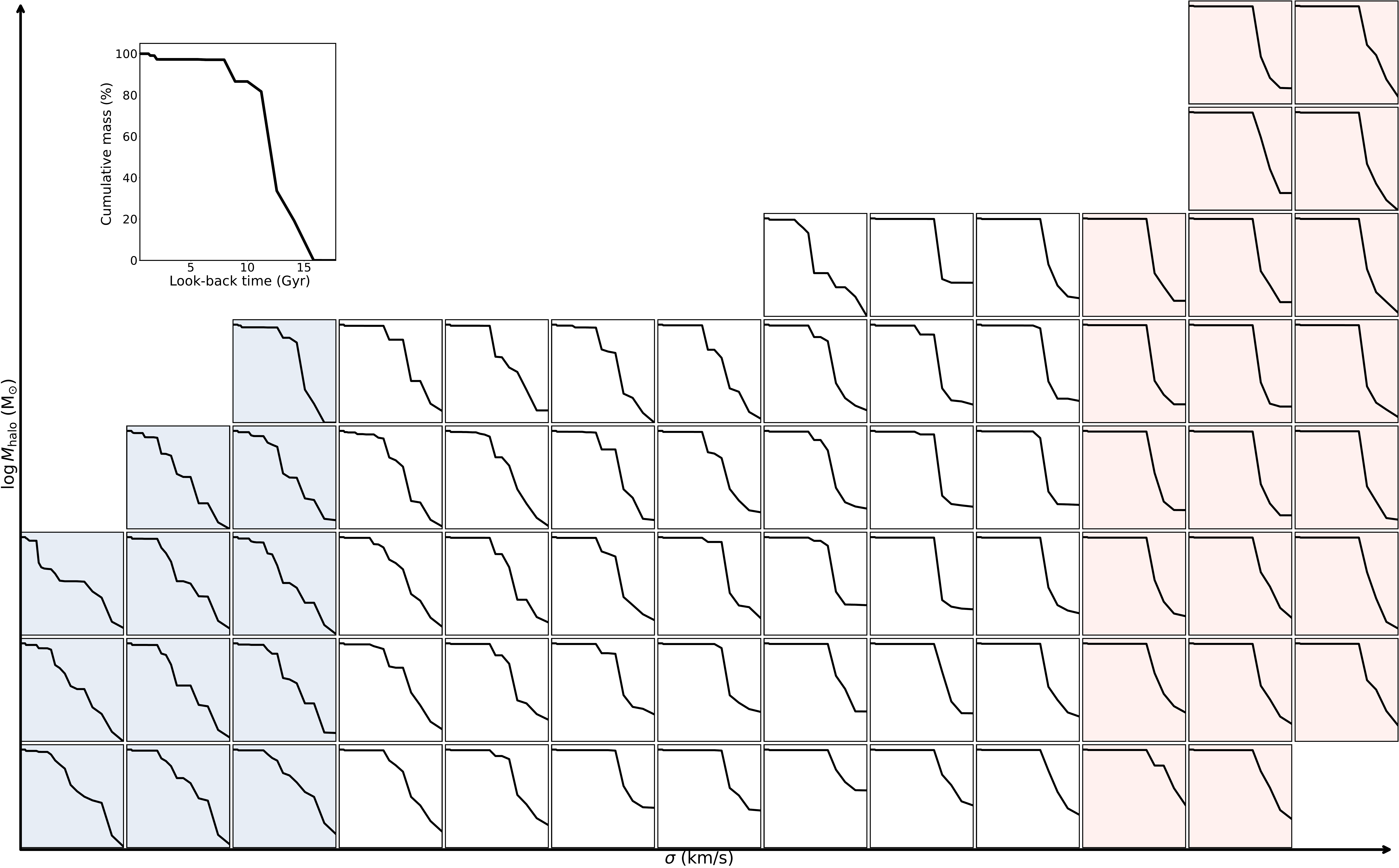}
    \caption{Star formation history for each central galaxy in the full sample,  across different ranges of $\sigma$ (horizontal arrow) and $M_{\rm halo}$ (vertical arrow). Each box represents the cumulative initial mass contribution, attributed to each SSP, as a function of the look-back time (in Gyr). The blue and red shadows highlight the different SFH of low-and-high-$\sigma$ central galaxies.}
    \label{fig:sfh}
\end{figure*}

The age of a galaxy reflects its entire SFH, but the numerical values alone cannot distinguish between a continuous, extended SFH and an old population followed by a quiescent phase with a recent burst. To address this issue, we present Figure~\ref{fig:sfh}, which shows the cumulative SFH for each bin of $\sigma$ (indicated by the horizontal arrow) and $M_{\rm halo}$ (vertical arrow) in the full sample. Each box represents the percentage of the cumulative mass of each SSP as a function of look-back time (in Gyr). The figure demonstrates that, for a fixed $\sigma$, the SFH becomes more extended as $M_{\rm halo}$ increases. Likewise, for a fixed $M_{\rm halo}$, the SFH is more extended for centrals with lower $\sigma$. Consequently, lower values of $\sigma$ and $M_{\rm halo}$ (lower left part of Figure~\ref{fig:sfh}) correspond to more extended SFHs, while the opposite holds true for the upper right part of the figure. In a study by \citet{2006MNRAS.366..499D}, which combines N-body simulations with semi-analytic techniques to investigate the formation and evolution of elliptical galaxies in a hierarchical merger model, they find that lower-mass elliptical galaxies exhibit extended star formation time-scales, consistent with our findings.

The similarity in the variations of Age and $Z$ with $\sigma$ and $M_{\rm halo}$ for both ICs and GCs indicates that the mechanisms by which centrals retain their hot gas halo (in the case of ICs) and accrete gas from the surroundings (in the case of GCs) are not fundamentally different. This implies that as gas cools down and flows into the galaxy, triggering the formation of new stars, there is an increase in $Z$ and a decrease in Age, regardless of whether the gas is retained from the halo or acquired through accretion.

\subsection{Probing the origin of the emission lines}
\label{gas}

In this section, we investigate the ISM properties by analysing $A_{\rm V}$ and the EWH$\alpha$, which reflect the galaxy's gas content. However, it is important to note that the mechanisms responsible for producing emission lines and interstellar extinction differ. While a galaxy showing the ionisation pattern typical of LINERs may exhibit signatures of H$\alpha$ and other spectral lines generated in the narrow line region near the galaxy nucleus, the effects of extinction can impact any region of the interstellar medium, regardless of the presence of nuclear activity.

\begin{figure*}
	\includegraphics[width=8.2cm]{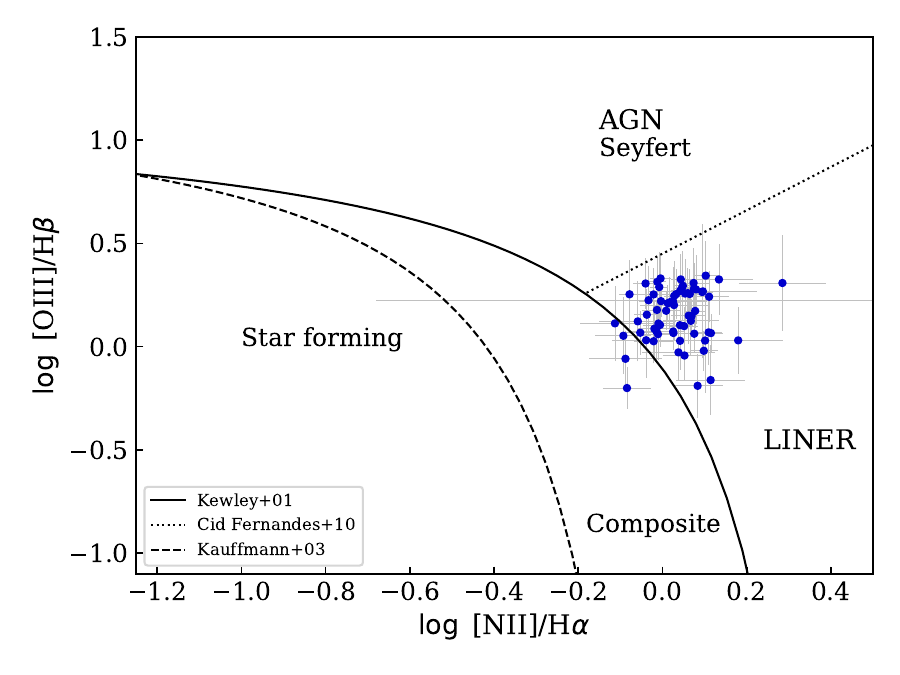}
    \includegraphics[width=8.2cm]{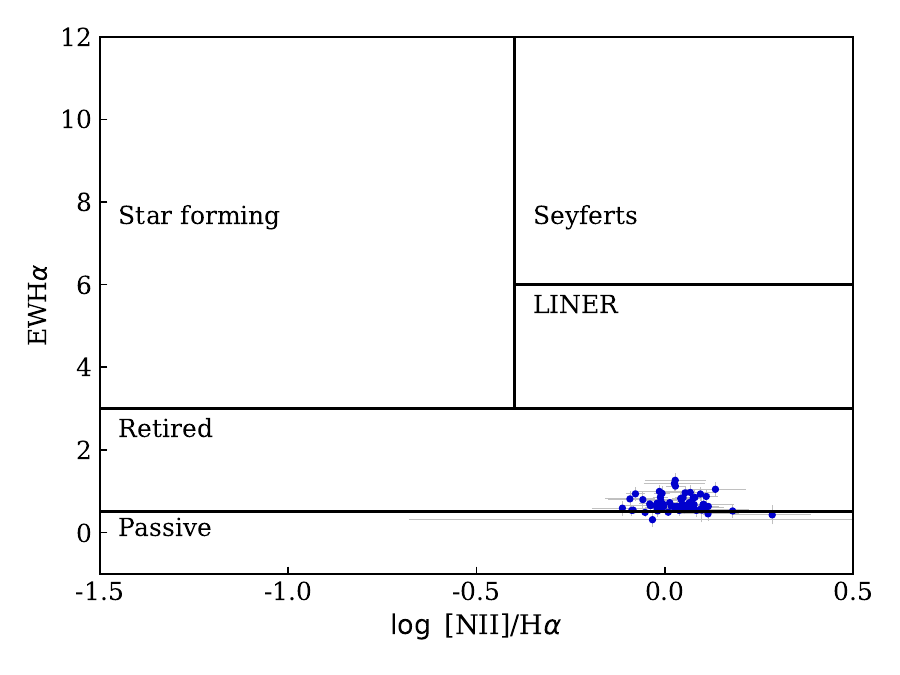}
    \caption{The left panel shows the BPT diagram, which reveals that the dominant ionisation pattern for central galaxies in the full sample is LINER, albeit with some cases exhibiting high uncertainty. In the right panel, the WHAN diagram indicates that HOLMES are the typical ionisation source for the central galaxies in our sample.}
    \label{fig:BPT_WHAN}
\end{figure*}

Regarding the gas phase of the ISM, variations in the EWH$\alpha$ can be attributed to different ionising agents such as young main-sequence stars, nuclear activity, and hot
evolved low-mass stars (HOLMES). To determine the ionising source in galaxies, we employ comparison techniques utilising the relative fluxes of emission lines. Elliptical galaxies, known for their low emission line intensities, are typically ionised by HOLMES \citep{1994A&A...292...13B,2010MNRAS.402.2187S,WHAN2}. We use the ``Baldwin, Phillips $\&$ Terlevich'' \citep[BPT]{Baldwin} to identify the ionisation source in our sample. The BPT diagram, shown in the left panel of Figure~\ref{fig:BPT_WHAN}, indicates that the dominant ionisation pattern of all studied galaxies is LINER. However, while commonly used, \citet{2008MNRAS.391L..29S} show that the LINER region in this diagram contains a combination of two distinct galaxy families: those hosting a weak AGN and those ionised by HOLMES.
Therefore, we employ the WHAN diagram \citep{Whan1} as a supplementary tool. As shown in the right panel of Fig.~\ref{fig:BPT_WHAN}, the WHAN diagram reveals that the typical ionisation source for central galaxies in our sample is HOLMES. Using both diagrams, we show that the emission line ratios and equivalent widths observed in our stacked spectra are fully consistent with that of ``retired'' galaxies, objects whose gas ionisation is due to HOLMES. 
It is important to note that each point in these diagrams represents the stacked data mentioned in Sect.~\ref{stacking_section}. Thus, even if individual galaxies within a stack exhibit intense emission lines due to nuclear activity or star formation, the resulting spectrum represents the characteristic ionisation pattern of the overall population within that specific stack. Our analysis confirms that HOLMES remain the primary ionisation source for central galaxies, even when considering ICs and GCs separately (see Figure~\ref{fig:BPT_WHAN_ICsGCs}). This is supported by the nearly identical diagnostic diagrams (BPT and WHAN) for these categories.

Our findings indicate that the ionising agent in the central galaxies of our sample is HOLMES, thereby eliminating the possibility of attributing variations in EWH$\alpha$ to a different source. Instead, we interpret these variations in EWH$\alpha$ in relation to $\sigma$ and $M_{\rm halo}$ as indicators of the mass fraction of ionisable gas within the galaxy. A study by \citet{Herpich} supports this interpretation, revealing that elliptical galaxies with EWH$\alpha$ values below $3\,$\AA\ share the same ionising agent, with differences between retired and passive galaxies primarily reflecting variations in the amount of ionisable gas present in the galaxy.

As far as $A_{\rm V}$ is concerned its numerical value depends on how the geometry of the grains and the specific efficiency of scattering affects the column density. If this column density is proportional to the column density of the ionised gas, it will also be proportional to EWH$\alpha$. Consequently, both $A_{\rm V}$ and EWH$\alpha$ will be indicative of the ISM presence. Their coefficients will be indicators of how much gas there is in the galaxy and also how these quantities vary according to $\sigma$ and $M_{\rm halo}$. Therefore, in the subsequent discussion, we will focus only on the results for EWH$\alpha$, which will indirectly apply to $A_{\rm V}$ as well. 

The observed decrease in EWH$\alpha$ as $\sigma$ increases (Fig.~\ref{fig:mh_sigma}, panels j, k and l) can be attributed to many mechanisms, including mergers, reabsorption of expelled gas by the galaxy itself, or absorption of peripheral gas. While this study does not explore all the possibilities exhaustively, we consider that central galaxies, unlike other ETGs, occupy the central regions of DM halos. As a result, they are more likely to be influenced by gas flows. Therefore, we propose that the differences in  H$\alpha$ emission arise from variations in the efficiency of gas accretion and cooling processes, which are influenced by a combination of cooling flows and feedback mechanisms.

The observed increase in EWH$\alpha$ with $M_{\rm halo}$ suggests a potential association with a higher cold gas infall rate. \citet{stott} propose an increased efficiency of the ICM cooling in massive halos to account for differences in the $L_X - T_X$ ratio between low and high $M_{\rm halo}$ systems. In order to provide a quantitative analysis, we develop a theoretical model that examines the interplay between two physical processes: AGN feedback and ICM cooling.

\subsection{Modelling the gas emission via ICM cooling flow and AGN feedback}
\label{ICMAGN}

Aiming to understand systematic differences in some properties of BCGs in a sample of 123 X-ray emitting clusters, \cite{stott}
investigated the relation between AGN feedback from the BCG and the ICM cooling. The authors argue that the thermodynamics of the ICM is set by
the relative importance of the energy $E_{\rm AGN}$ released by AGN feedback over the cluster history to the thermal energy $E_{\rm ICM}$ of the ICM.
An estimate of $E_{\rm AGN}$ was obtained for a typical radio-loud galaxy in their sample and expressed as a function of the feedback efficiency $\eta$, while considerations involving the dynamical equilibrium and the ICM temperature were then used to estimate the total thermal energy of the intracluster medium as a function of $ M_{500} $\footnote{The mass contained in a radius within which the density of matter is 500 times the mean mass density of the Universe.} and the ICM temperature $ T_{X} $  as

\begin{equation}
E_{\rm ICM} = \frac{3k_B}{2\mu m_p}T_{X}fM_{500},
\end{equation}

\noindent where $ f $ is the ICM mass fraction, $ k_{B} $ the Boltzmann constant, $ \mu $ is the mean molecular weight of the constituent particles of the ICM and $ m_p $ is the proton's mass.

If the presence of ionisable gas in a central galaxy is due to ICM cooling, the parameterisation of EWH$\alpha$ we have performed in Sect.~\ref{Parameterisation_with_sigma_and_halo_mass} as a function of $\sigma$ and $M_{\rm halo}$ maps the ICM cooling efficiency as compared to the energy output due to AGN feedback on the ICM. The typical timescales for ICM cooling in the centre of clusters are much lower than $1\,$Gyr, and so we expect that, for a fixed total thermal energy for the ICM, the occurrence of an emission line in a central galaxy could be therefore the result of the interplay between the energy dissipation of the ICM by X-ray emission and the instantaneous kinetic energy output from AGN feedback, irrespective of the total energy that the latter has released to the ICM over cosmic time. In this case, the ratio $ \xi $ between the instantaneous kinetic power of the AGN feedback ($PE_{\rm BH}$) and the thermal energy of the ICM, ($E_{\rm ICM}$), i.e.

\begin{equation}
\xi = \frac{PE_{\rm  BH}}{E_{\rm  ICM}},
\label{eq:xi}
\end{equation}

\noindent effectively defines the H$\alpha$ emission in central galaxies. Notice that $\xi$ is equivalent (except for a multiplicative constant) to the $E_{\rm AGN}/E_{\rm ICM}$ ratio proposed by  \cite{stott} for a fixed AGN lifetime.

In order to model the EWH$\alpha$ observed in our sample of  central galaxies, we expand the above prescription, re-writing equation \ref{eq:xi} as an explicit function of $\sigma$ and $M_{\rm halo}$.
We assume that the total mass $M_{\rm halo}$ of the cluster scales as $M_{\rm halo} \propto M_{500}$ and is also proportional to the cluster total volume, so that ${M_{\rm halo}} \propto R^3$, where $R$ is a characteristic radius. We also assume an equilibrium condition where the temperature of the gas, $T_X$, scales with the average kinetic energy of the galaxies in the cluster, i.e. to the square of the dispersion in the values of the individual velocities of the galaxies of the system \citep{1993ApJ...415L..17L}. For a spherically symmetric, homogeneous and virialised distribution of particles of total mass $M$ and radius $R$, $\sigma^2$ scales with $M/R$. We then obtain

\begin{equation}
E_{\rm ICM} \propto \frac{M_{\rm halo}}{R} M_{\rm halo} \propto \frac{M_{\rm halo}}{M_{\rm halo}^{1/3}}M_{\rm halo}  \propto M_{\rm halo}^{1.67}.
\end{equation}

\noindent We further assume that the instantaneous kinetic power of the AGN feedback $ PE_{\rm BH} $ scales with the accretion rate by the SMBH with a constant efficiency $ \varepsilon $, i.e.

\begin{equation}
\centering
\label{eq:power_accretion}
    PE_{\rm BH} = \varepsilon \dot{M_{\rm BH}} c^2.
\end{equation}

\noindent We now need a recipe for expressing the accretion rate of the SMBH in terms of the physical parameters that describe the galaxies in each stack, i.e. their $\sigma$ and $M_{\rm halo}$. 
The Bondi \citep{1952MNRAS.112..195B} mechanism has been shown to be efficient in describing several properties of BCGs \citep{2014arXiv1406.6366F}. In this scenario, the feeding of the SMBH occurs through the accretion of hot gas from its surroundings, when the gravitational potential exceeds the thermal energy of the gas. This accretion mode is radiatively inefficient, producing a feedback that is mostly mechanical, in the form of jets. The fact that radio galaxies present typically an early-type morphology, and that the probability of an ETG to be a radio galaxy is even higher if it is a central system, reinforces the use of the Bondi scenario as a first approximation.
This model implies that the accretion rate of the SMBH is proportional to the square of its mass. Assuming that the gas densities, the speed of sound in the surrounding medium and the Bondi radii are homogeneous across all bins of $\sigma$ and $M_{\rm halo}$, we get $\dot{M}_{\rm BH} \propto M^2_{\rm BH}$, what results in

\begin{equation}
    PE_{\rm BH} \propto M^2_{\rm BH}.
\end{equation}

\noindent To explicitly couple $ PE_{\rm BH} $ with the global properties of their host galaxies, we use the parameters of the $M_{\rm BH} - \sigma $ relation derived by \cite{Kormendy}. This results in  $M_{\rm BH} \propto \sigma^{4.38}$, and therefore

\begin{equation}
\label{eq:power_sigma}
    PE_{\rm BH} \propto \sigma^{8.76}. 
\end{equation}

\noindent Substituting the expressions for $ PE_{\rm BH} $ and $ E_{\rm ICM} $ in  equation \ref{eq:xi}, we finally obtain

\begin{equation}
    \xi \propto \sigma^{8.76} M_{\rm halo}^{-1.67}.
\end{equation}

\noindent Low values of $\xi$ imply that the power of the AGN feedback is small for a fixed total thermal energy of the ICM. This in turn results in large ICM cooling rates; the flow of gas into the central galaxy then enhances  H$\alpha$ emission. The effect of the ICM cooling in the H$\alpha$ luminosity can therefore be parameterised as
\begin{equation}
\label{eq:L_alpha}
    L_{{\rm H}\alpha} \propto \xi^{\gamma},
\end{equation}

\noindent where $ \gamma $ is a \emph{negative} coefficient whose absolute value depends on the detailed physics of ICM cooling and heating mechanisms.

The above prescription for $L_{{\rm H}\alpha}$ can be transformed into a similar one for ${\rm{EW}H}\alpha$ in the following way. By definition,

\begin{equation}
    {\rm{EWH}}\alpha=\frac{F_{{\rm H}\alpha}}{\overline{F_{\rm C}}},
\end{equation}

\noindent where $F_{\rm H\alpha}$ is the flux of the H$\alpha$ line and $\overline{F_{\rm C}}$ is the average level of the spectral continuum at the position of the H$\alpha$ line. This expression can be converted in a ratio involving $L_{\rm H\alpha}$ by means of

\begin{equation}
    {\rm{EWH}}\alpha=\frac{F_{{\rm H}\alpha}}{\overline{F_{\rm C}}}\frac{4\pi D^2}{4\pi D^2}\equiv\frac{L_{{\rm H}\alpha}}{L_{\rm C}},
\end{equation}

\noindent where $D$ is the luminosity distance of the galaxy. The factor $L_{\rm C}\equiv \overline{F_{\rm C}}4\pi D^2$ -- the average luminosity density of the spectral continuum at the position of the H$\alpha$ line -- can be estimated by the (fibre) luminosity of the galaxy in a spectral window around the H$\alpha$ line, what roughly corresponds to the high wavelength limit of the passband of the SDSS $r$ filter at the typical redshifts of our sample of centrals. The $r$-band absolute magnitudes are available from the SDSS, but this is not a good proxy for $L_{\rm C}$, because in general only a fraction of the galaxy light is included in the SDSS spectroscopic fibre (and this fraction is strongly dependent on the redshift of the galaxy).
Assuming that the galaxy is homogeneous and therefore  the ratio between the fibre ($f_{r,F}$) and total ($f_{r,T}$) $r$-band fluxes of the galaxy is equal to the ratio between the respective luminosities ($L_{r,F}$, $L_{r,T}$), we get

\begin{equation}
    L_{r,F}=\frac{f_{r,F}}{f_{r,T}}L_{r,T}.
\end{equation}

\noindent Using the CASJOBS\footnote{https://skyserver.sdss.org/casjobs/} environment, we obtain the values of $f_{r,F}$,  $f_{r,T}$ and $L_{r,F}$ for the galaxies in each stack and derive $L_{r,F}$. Finally, we average the $L_{r,F}$ values across all galaxies in a stack to get $L_{\rm C}$.

\begin{figure*}
\centering
\captionsetup{justification=centering}
\includegraphics[width=16cm]{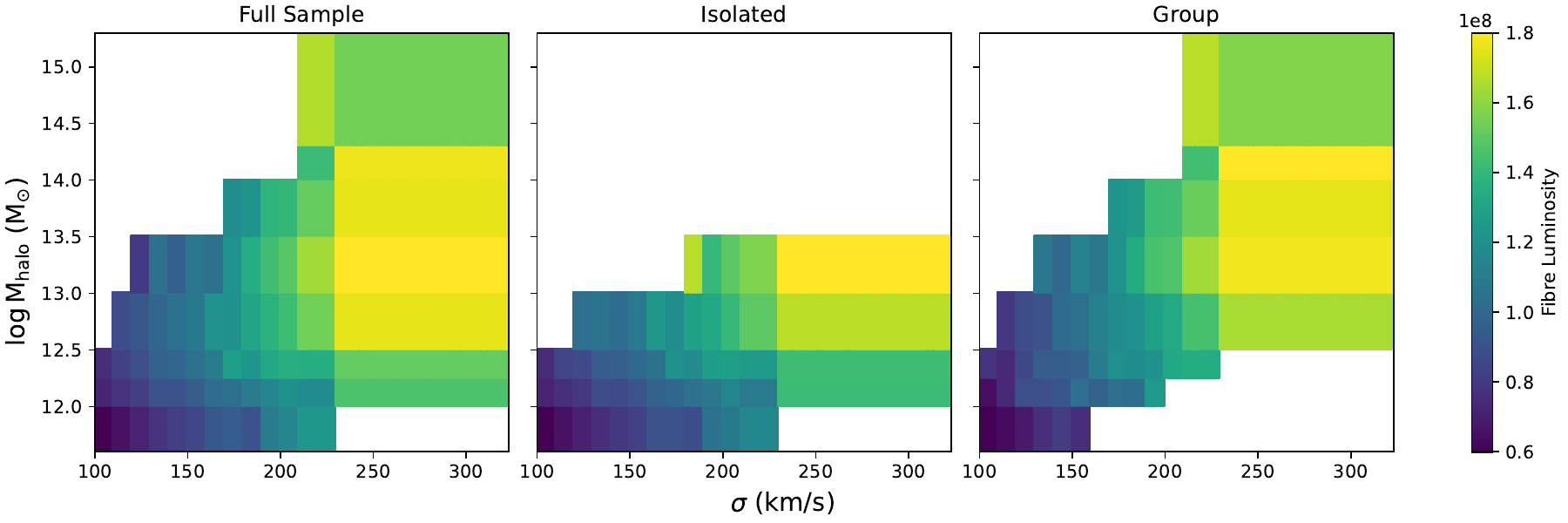}
\caption{Following the same pattern of Fig.~\ref{fig:resvaz}, we present the correlation of the fibre luminosity with $\sigma$ and $M_{\rm halo}$ for each central galaxy bin in the full sample, ICs and GCs, from left to right, respectively.}
\label{fig:mean_model}
\end{figure*}

It is important to note that $L_{\rm C}$ is the average $r$-band fibre luminosity per stack; just like other physical parameters, it may also be a function of $\sigma$ and $M_{\rm halo}$.
We have quantified such dependence in Figure~\ref{fig:mean_model},  using the same fitting procedure described Sect.~\ref{Parameterisation_with_sigma_and_halo_mass}. The best-fit coefficients are obtained using a non-linear method of least squares using the curve fit from SciPy Python library. The uncertainties applied to this method are given by the standard deviation measured for each property for the individual galaxies in each stack.
After performing these fits, the EWH$\alpha$ can finally be expressed as

\begin{equation}
\label{eq:coef_all}
\begin{split}
\log \rm{EWH}\alpha & \propto [8.76\gamma - (0.75 \pm 0.04)]\log \sigma  + \\ 
 & [-1.67\gamma - (0.042 \pm 0.006)]\log M_{\rm halo}
\end{split}
\end{equation}

\noindent for the full sample, and 

\begin{equation}
\label{eq:coefsub_iso}
\begin{split}
\log \rm{EWH}\alpha \propto & [8.76\gamma - (0.66\pm0.03)]\log \sigma  + \\ 
 & [-1.67\gamma - (0.153\pm0.008)]\log M_{\rm halo} \\
\end{split}
\end{equation}

\noindent and

\begin{equation}
\label{eq:coefsub_clu}
\begin{split}
\log \rm{EWH}\alpha \propto & [8.76\gamma - (0.86\pm0.04)]\log \sigma  + \\ 
 & [-1.67\gamma - (0.030\pm0.005)]\log M_{\rm halo}, \\
\end{split}
\end{equation}

\noindent respectively, for ICs and GCs.

Equations \ref{eq:coef_all} to \ref{eq:coefsub_clu} contain a set of theoretical coefficients of the relation ${\rm EWH}\alpha \sim \sigma^A M_{\rm halo}^\textit{B}$ that can be directly compared to those obtained
in Sect.~\ref{Parameterisation_with_sigma_and_halo_mass}. Such coefficients are numerically defined except for the exponent $\gamma$, which is an unknown in our model but whose value can be obtained directly by comparison with the measured coefficients in Table~\ref{tab:coef_table}. It is important to note that, as $\gamma$ is included in both theoretical coefficients, it can be solved for the value that produces the best agreement between the theoretical and measured ones. 
We obtain $\gamma_{\rm full} = -0.05\pm0.03$, $\gamma_{\rm iso} = -0.16\pm0.07$ and $\gamma_{\rm group} = -0.01\pm0.01$ for the full sample, ICs and GCs, respectively. Such values result in theoretical coefficients that are consistent with the measured ones (Tab.~\ref{tab:coef_table}) within $2\sigma$.
The existence of a solution for $\gamma$ that allows for a good match between the theoretical and measured coefficients -- and also presents the correct (negative) sign as expected -- is suggestive that the AGN feedback power relative to the ICM internal energy indeed regulates the gas flow into the body of the central galaxy.

Isolated centrals are the population for which we obtain the highest absolute value of $\gamma$; taking into account its uncertainty, $\gamma_{\rm iso}$ is different from zero at more than $95$\% statistical confidence. This suggests that, for this sub-population, EWH$\alpha$ responds to the ratio between $PE_{\rm BH}$ and $E_{\rm ICM}$. The interpretation given earlier to explain the observed trends for the stellar populations for IC still holds: ICs are able to retain their halo of hot gas more efficiently than GCs, which often lose a considerable amount of their gas through environmental processes. By maintaining their halo of hot gas, ICs can continue cooling, forming new stars and replenishing their ionised gas budget, eventually triggering an active phase and repeating this process as long as the gas reservoir exists. This leads to a more extended SFH, a younger and more metal-rich stellar population, and an ionised gas signature regulated by AGN feedback.

 
On the other hand, the value of $\gamma_{\rm group}$ is consistent with zero within its uncertainty. If  $\gamma$ is set to zero, then $L_{{\rm H}\alpha}$ is independent of the ratio between the kinetic power of the AGN  and the thermal energy of the ICM and our model fails to adequately describe the EWH$\alpha$ values.
This suggests that the GCs are less efficient in retaining their hot gas halo compared to ICs and the signal of the interplay between such processes is confounded by other mechanisms that also participate in the heating and cooling of the ICM in GCs. In fact, gas sloshing due to galaxy mergers or simply by the motions of galaxies close to the central regions of the halo -- e.g. \cite{2011MNRAS.414.1493R} -- may also be relevant sources of ICM heating in the group environment. Conversely, central galaxies in groups are subjected to merging, a phenomenon that can displace the central's SMBH and effectively disrupt the feeding-feedback cycle \citep[e.g.][]{2023MNRAS.522..948C}.
The scenario of accretion of gas-rich, low-mass galaxies that we have already invoked to explain the SFH of GCs, may also play an important role in setting the H$\alpha$ emission in such galaxies, as it introduces both younger stars and gas. Considering the mass fractions in central galaxies with low $\sigma$ values shown in Figure~\ref{fig:sfh}, it is possible that a significant number of stars have recently been included into the system via mergers; the recently-acquired gas is then ionised by the radiation field of HOLMES, contributing to the EWH$\alpha$ signature observed in our sample. In short, the gas deposition in central galaxies in groups seems to be more complex than a single AGN power versus ICM thermal energy prescription, as opposed to ICs; instead, other mechanisms that also
contribute to ICM heating and cooling processes must play an important role.

\section{Summary}
\label{sec:Summary}

In this study, we investigate the stellar population and interstellar medium properties of a sample of $15,107$ early-type central galaxies from the SPIDER survey \citep{spiderviii}. The aim is to understand the evolution of the baryonic content in these galaxies. Using optical spectra obtained from SDSS, we derive key parameters including Age, $Z$, $A_{\rm V}$, and EWH$\alpha$. Our analysis focuses on examining the dependence of these properties on central $\sigma$ and $M_{\rm halo}$ for both isolated and group centrals. To identify ionisation sources, we employ BPT and WHAN diagnostic diagrams and propose a scenario based on a Bondi accretion regime to describe the gas properties in central galaxies. The summarised findings of our study are as follows: 

\begin{itemize}
    \item We find that ICs and GCs show a similar trend with $\sigma$ -- older and more metal rich stellar populations as $\sigma$ increases. Additionally, when considering a fixed $\sigma$, we observe a trend of younger Ages associated with higher values of $M_{\rm halo}$s. Remarkably, both ICs and GCs display consistent correlations between these parameters, suggesting a coherent final star formation history. This holds true regardless of whether the central galaxy acquires gas from its own halo (as in ICs) or from accreted gas-rich systems (as in GCs).

    \item In both ICs and full sample, we observe a consistent increase in $A_{\rm V}$ and EWH$\alpha$ as $\sigma$ decreases and $M_{\rm halo}$ increases. The trends of these parameters with $M_{\rm halo}$ are not clear for CGs due to larger scatter. For ICs, the variation with $M_{\rm halo}$ becomes less evident at $\sigma > 200\,\rm km\,s^{-1}$.  Also, the variation with sigma flattens out for all systems above $200\,\rm km\,s^{-1}$.
   The similarity between these properties is further supported by the values obtained through our parameterisation, indicating that both $\sigma$ and $M_{\rm halo}$ serve as indicators of the ionisable gas content in the central galaxies of our sample. The linear combination of these parameters provides valuable insights into the properties of the ionisable gas in these systems.

    \item In the case of ICs, we successfully reproduce the observed values of EWH$\alpha$ by utilising a model that takes into account the interplay between the kinetic power of AGN and the thermal energy of the ICM, providing an explanation for gas deposition in these central galaxies. However, when applying the same model to GCs, it fails to adequately describe the EWH$\alpha$ values. These findings suggest that ICs are more efficient in retaining their hot gas halo compared to GCs. In the case of GCs, the influence of AGN feedback on ICM thermodynamics may be confounded by other mechanisms that also contribute to the heating and cooling processes within the ICM in GCs.
\end{itemize}

Our findings indicate intrinsic differences between isolated and group central galaxies. Isolated central galaxies effectively retain their hot gas halo, enabling efficient cooling and reheating. In contrast, gas deposition in group central galaxies is a complex process, involving factors beyond the direct interaction between AGN power and the thermal energy of the intra-cluster medium. These observations suggest that the mechanisms governing gas dynamics in isolated and group central galaxies are likely influenced by additional factors.

\section*{Acknowledgements}

VL acknowledges the Coordenação de Aperfeiçoamento de Pessoal de Nível Superior -- CAPES scholarship
through the grants 88882.427898/2019-01.
SBR acknowledges support from Conselho
Nacional de Desenvolvimento Científico e Tecnológico -- CNPq.
RRdC aknowledges the financial support from FAPESP through the grant $2020/15245-2$.

Funding for SDSS-III has been provided by the Alfred P. Sloan Foundation, the Participating Institutions, the National Science Foundation, and the U.S. Department of Energy Office of Science. The SDSS-III web site is http://www.sdss3.org/. SDSS-III is managed by the Astrophysical Research Consortium for the Participating Institutions of the SDSS-III Collaboration including the University of Arizona, the Brazilian Participation Group, Brookhaven National Laboratory, Carnegie Mellon University, University of Florida, the French Participation Group, the German Participation Group, Harvard University, the Instituto de Astrofisica de Canarias, the Michigan State/Notre Dame/JINA Participation Group, Johns Hopkins University, Lawrence Berkeley National Laboratory, Max Planck Institute for Astrophysics, Max Planck Institute for Extraterrestrial Physics, New Mexico State University, New York University, Ohio State University, Pennsylvania State University, University of Portsmouth, Princeton University, the Spanish Participation Group, University of Tokyo, University of Utah, Vanderbilt University, University of Virginia, University of Washington, and Yale University.

This work was made possible thanks to the open-source software packages AstroPy \citep{2018AJ....156..123A}, Matplotib \citep{Hunter:2007}, NumPy \citep{5725236}, Pandas \citep{mckinney-proc-scipy-2010} and SciPy \citep{2020SciPy-NMeth}.


\section*{Data Availability}

All data used in this article are publicly available at https://www.sdss3.org/.
 



\bibliographystyle{mnras}
\bibliography{Lorenzoni_23} 




\appendix
\section{}
\label{appendix1}

Tables~\ref{ngal} to \ref{tab:ngal_group} present the numbers of central galaxies per $\sigma$ and $M_{\rm halo}$ bin, along with the S/N (in parentheses) for the full sample, isolated, and group centrals respectively.

\begin{figure*}
	\includegraphics[width=\textwidth]{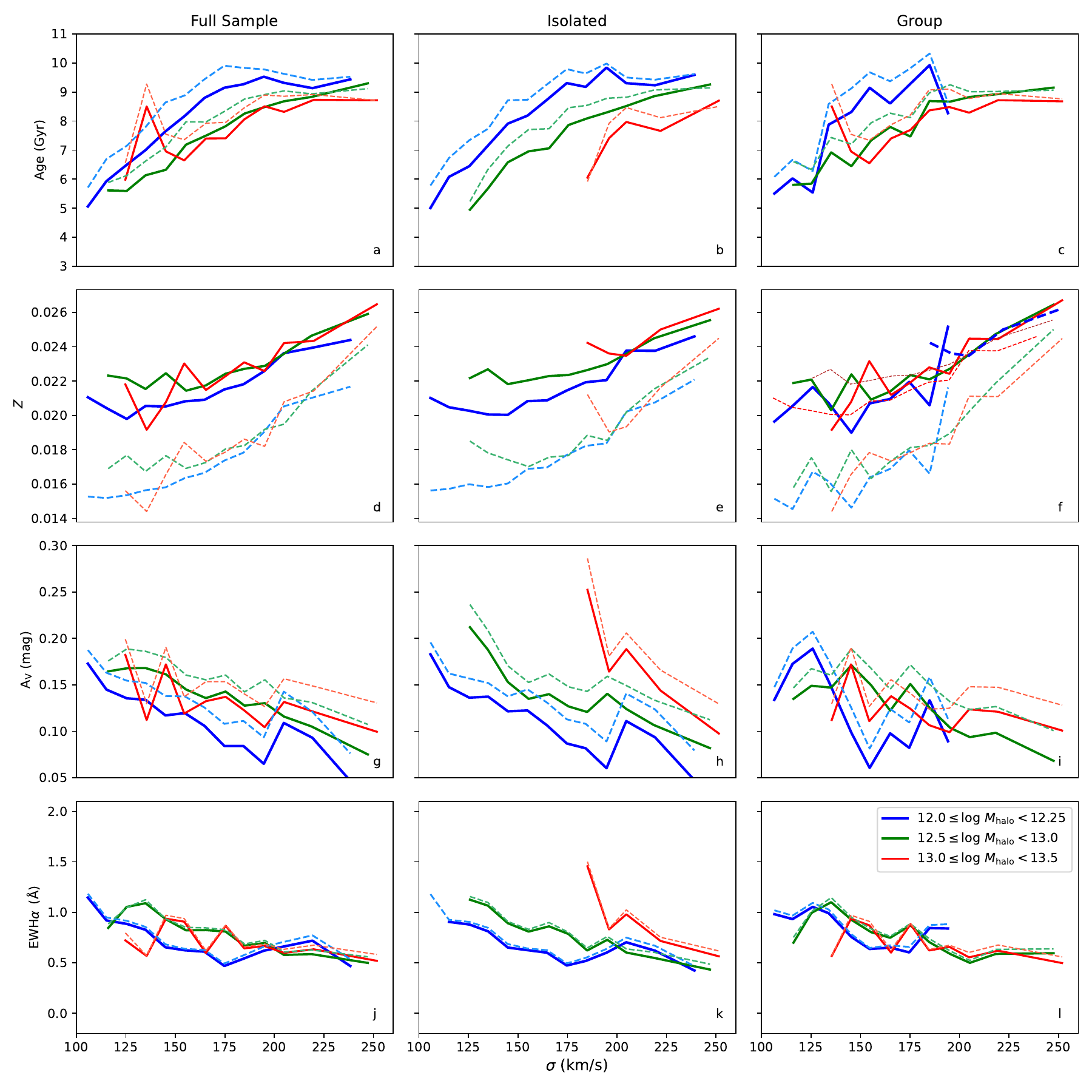}
    \caption{Relationship between Age, $Z$, $A_{\rm V}$, and EWH$\alpha$ with $\sigma$, represented from top to bottom, respectively, for GM templates (solid lines) in comparison to MILES models (dashed lines). The full sample, along with ICs and GCs, is presented from left to right, respectively. Additionally, we included three $M_{\rm halo}$ ranges, distinguished by colours.}
    \label{fig:mh_sigma_GM}
\end{figure*}

In Figure~\ref{fig:mh_sigma_GM}, we compare the results from Figure~\ref{fig:mh_sigma} (represented by dashed lines) to the results obtained using GM templates (solid lines). While the parameter values vary as expected, the trends with $\sigma$ and $M_{\rm halo}$ are very similar.

\begin{figure*}
	\includegraphics[width=8.3cm]{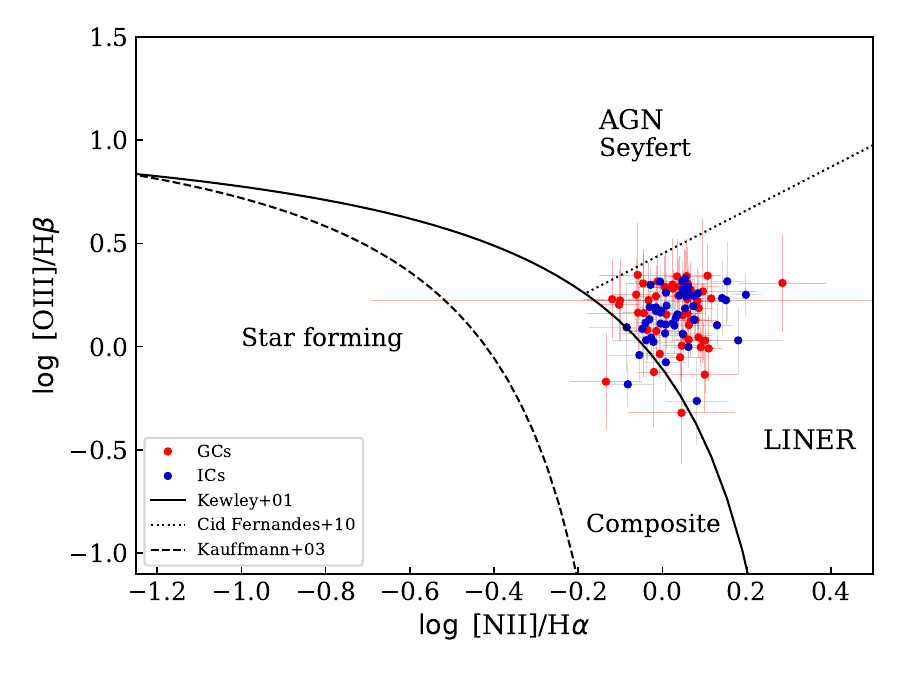}
    \includegraphics[width=8.3cm]{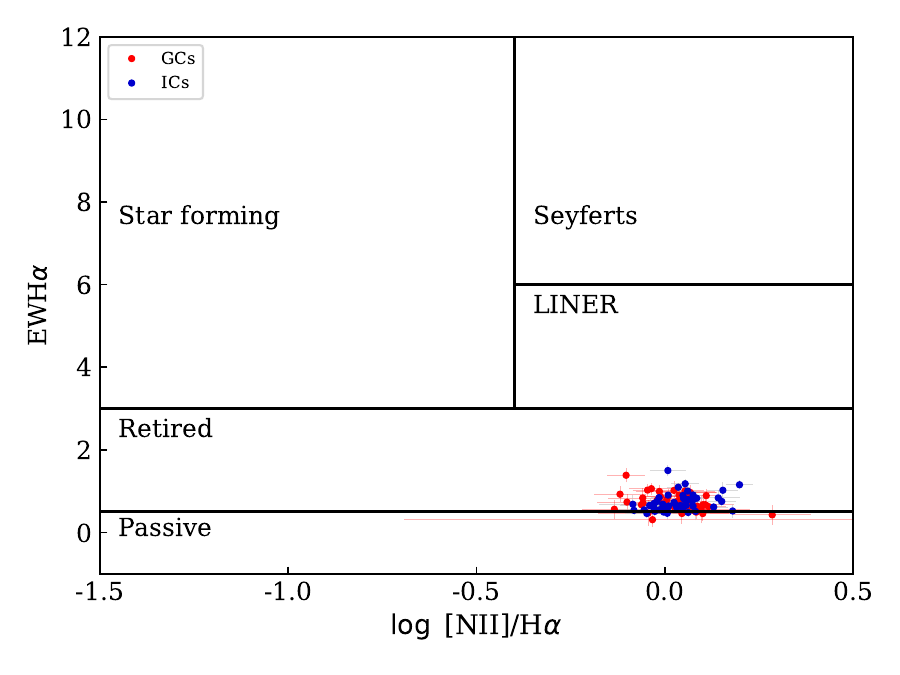}
    \caption{The left panel shows the BPT diagram for ICs (blue dots) and GCs (red dots), which reveals that the dominant ionisation pattern for central galaxies in both sub-samples is LINER, albeit with some cases exhibiting high uncertainty. In the right panel, the WHAN diagram indicates that HOLMES are the typical ionisation source for ICs and GCs.}
    \label{fig:BPT_WHAN_ICsGCs}
\end{figure*}

Figure~\ref{fig:BPT_WHAN_ICsGCs} presents the BPT and WHAN diagnostic diagrams for isolated (blue dots) and group (red dots) centrals. For both the BPT and WHAN diagram, the results for both samples remain consistent with Figure~\ref{fig:BPT_WHAN}.

\begin{table*}
	\centering
	\caption{Number of Galaxies (Signal-to-noise ratio) per bin of  $\sigma$ and $M_{\rm halo}$ in the full sample. }
	\label{ngal}
	\begin{tabular}{ccccccccc} 
	\hline
		& \multicolumn{8}{c}{$\log M_{\rm halo}/\rm{M_\odot}$} \\
		\hline
 & 11.60 &  12.00 & 12.25 & 12.50 &  13.00 & 13.50 & 14.00 & 14.30 \\
$\sigma$ (km s$^{-1}$) & 12.00 &  12.25 & 12.50 & 13.00 & 13.50 & 14.00 & 14.30 & 15.30 \\ %
		\hline
        100-110 & 318 (289)  & 167 (220) & 34 (107)  &   &   &   &   &   \\
        110-120 & 458 (355)  & 359 (336) & 105 (186)  &   24 (96)  &  &  &  & \\
        120-130 & 506 (392)  & 499 (396) & 168 (245)  &  40 (127) &   5 (74)    &   &  &   \\
        130-140 & 467 (380)  & 679 (476) & 324 (350) &  98 (192)  &   6 (59)   &   &  &    \\
        140-150 & 334 (330) & 639 (483) & 365 (385) &  145 (258) &  9 (71) &   &  &  \\
        150-160 & 248 (305)  & 617 (497) & 520 (459)  &  243 (319) &   21 (101)  &  &  &  \\
        160-170 & 126 (220) & 451 (435) & 468 (461) &  336 (395) &   37 (136)    &  &  &  \\
        170-180 & 69 (169)  & 293 (359) & 428 (451) &  374 (427) &   50 (165)    &   9 (86) &  &  \\
        180-190 & 45 (142)  & 212 (311) & 344 (421)  &  445 (474)  &   83 (219)  &   6 (77)  &   &  \\
        190-200 & 20 (102) & 104 (217) & 215 (331)  &  380 (449) &   91 (232)   &   23 (107)  &    & \\
        200-210 &  13 (83) & 60 (170) & 162 (313) &  389 (465) &  151 (311)   &   28 (146) &     &  \\
        210-230 &  14 (98) & 42 (149) & 114 (267) &  448 (538) &   306 (460)  &   78 (241) &      6 (82)   &  6 (93)  \\
        230-323 &      & 24 (134) & 30 (147) &  301 (458) &   383 (554)  &   231 (452) &      51 (220)  & 26 (159) \\
        \hline
\end{tabular}
\end{table*}

\begin{table*}
	\centering
	\caption{Number of Galaxies (Signal-to-noise ratio) per bin of  $\sigma$ and $M_{\rm halo}$ for the isolated centrals.}
	\label{tab:ngal_isolated}
	\begin{tabular}{ccccccccc} 
	\hline
		& \multicolumn{8}{c}{$\log M_{\rm halo}/\rm{M_\odot}$} \\
		\hline
 & 11.60 &  12.00 & 12.25 & 12.50 &  13.00 & 13.50 & 14.00 & 14.30 \\
$\sigma$ (km s$^{-1}$) & 12.00 &  12.25 & 12.50 & 13.00 & 13.50 & 14.00 & 14.30 & 15.30 \\ %
		\hline

		\hline
        100-110 & 292 (276) & 148 (208) &  18 (80) &            &          &     &  &   \\ 
        110-120 & 423 (339) & 329 (324) &  81 (165) &           &          &     &  &   \\
        120-130 & 458 (373) & 454 (379) & 128 (211) &  14 (80)  &          &     &  &   \\ 
        130-140 & 432 (364) & 616 (451) & 244 (300) &  50 (152) &          &     &  &   \\ 
        140-150 & 303 (313) & 564 (455) & 312 (358) &  67 (178) &          &     &  &   \\ 
        150-160 & 221 (289) & 569 (475) & 425 (415) & 128 (228) &          &     &  &   \\
        160-170 & 115 (210) & 412 (414) & 384 (415) & 182 (289) &          &     &  &   \\ 
        170-180 &  65 (164) & 257 (333) & 356 (412) & 207 (311) &          &     &  &   \\ 
        180-190 &  41 (135) & 183 (288) & 283 (381) & 269 (376) &  5 (73)  &     &  &   \\ 
        190-200 &  16 (92) &  93 (204) & 172 (295) & 241 (356)  &  6 (90)  &     &  &   \\ 
        200-210 &  13 (83) &  56 (164) & 123 (272) & 217 (351)  & 13 (99)  &     &  &   \\ 
        210-230 &  14 (98) &  39 (142) &  91 (238) & 250 (393)  & 33 (157) &     &  &   \\
        230-323 &          &  20 (123) &  23 (129) & 182 (350)  & 56 (201) &     &  &   \\
        \hline
\end{tabular}
\end{table*}

\begin{table*}
	\centering
	\caption{Number of Galaxies (Signal-to-noise ratio) per bin of  $\sigma$ and $M_{\rm halo}$ for the group centrals.}
	\label{tab:ngal_group}
	\begin{tabular}{ccccccccc} 
	\hline
		& \multicolumn{8}{c}{$\log M_{\rm halo}/\rm{M_\odot}$} \\
		\hline
 & 11.60 &  12.00 & 12.25 & 12.50 &  13.00 & 13.50 & 14.00 & 14.30 \\
$\sigma$ (km s$^{-1}$) & 12.00 &  12.25 & 12.50 & 13.00 & 13.50 & 14.00 & 14.30 & 15.30 \\ %
		\hline
        100-110 & 14 (67) & 10 (66) & 12 (67) &        &        &        &    &     \\  
        110-120 & 15 (82) & 18 (80) & 20 (84) & 18 (86)  &        &        &    &     \\   
        120-130 & 16 (81) & 24 (95) & 33 (118) & 26 (105)  &        &        &    &     \\   
        130-140 & 13 (78) & 38 (125) & 67 (172) & 45 (123)  &   6 (59) &        &    &     \\   
        140-150 & 17 (92) & 38 (114) & 42 (129) & 70 (179)  &   9 (71) &        &    &     \\   
        150-160 & 16 (81) & 31 (121) & 71 (171) & 109 (222) &  20 (98) &        &    &     \\   
        160-170 &       & 28 (120) & 66 (183) & 142 (263) &  36 (134) &        &    &     \\   
        170-180 &       & 24 (118) & 50 (161) & 151 (283) &  45 (155) &  9 (86)  &    &    \\  
        180-190 &       & 15 (92) & 44 (160) & 160 (282) &  74 (208) &  6 (77)  &    &    \\  
        190-200 &       & 9 (86)  & 36 (147) & 122 (259) &  84 (220) & 21 (101)  &    &    \\  
        200-210 &       &       & 29 (142) & 156 (295) & 132 (291) & 27 (143)  &    &    \\  
        210-230 &       &       & 19 (117) & 175 (353) & 262 (425) & 73 (235)  &  6 (82) & 6 (93)   \\  
        230-323 &       &       &       & 105 (287) & 308 (503) & 218 (437) & 49 (217) & 24 (156)  \\ 
        \hline
\end{tabular}
\end{table*}

\bsp	
\label{lastpage}
\end{document}